\documentclass{tCPH2e}

\usepackage{epstopdf}
\usepackage{subfigure}
\usepackage{color}
\usepackage[breaklinks=true,colorlinks=true,linkcolor=blue,urlcolor=blue,citecolor=blue]{hyperref}
\def\bra#1{{\left\langle #1 \right|}}
\def\ket#1{{\left| #1 \right\rangle}}
\def\expect#1{\left\langle #1 \right\rangle}
\newcommand{\identity}{\leavevmode\hbox{\small1\kern-3.2pt\normalsize1}}

\begin{document}


\title{Hybrid quantum computing with ancillas}

\author{
\name{Timothy J. Proctor\textsuperscript{a,b}
and Viv Kendon\textsuperscript{c}$^{\ast}$\thanks{$^\ast$Corresponding author. Email: viv.kendon@durham.ac.uk}}
\affil{\textsuperscript{a}School of Physics and Astronomy, E C Stoner Building, University of Leeds, Leeds, LS2 9JT, UK; \textsuperscript{b}Berkeley Quantum Information and Computation Center, Department of Chemistry, University of California, Berkeley, CA 94720, USA; \textsuperscript{c}Department of Physics, Durham University, South Road, Durham, DH1 3LE, UK}
}

\maketitle

\begin{abstract}
In the quest to build a practical quantum computer, it is important to use efficient schemes for enacting the elementary quantum operations from which quantum computer programs are constructed. The opposing requirements of well-protected quantum data and fast quantum operations must be balanced to maintain the integrity of the quantum information throughout the computation. One important approach to quantum operations is to use an extra quantum system -- an ancilla -- to interact with the quantum data register. Ancillas can mediate interactions between separated quantum registers, and by using fresh ancillas for each quantum operation, data integrity can be preserved for longer. This review provides an overview of the basic concepts of the gate model quantum computer architecture, including the different possible forms of information encodings -- from base two up to continuous variables -- and a more detailed description of how the main types of ancilla-mediated quantum operations provide efficient quantum gates.
\end{abstract}


\section{Introduction}
\begin{figure}[h!]
\begin{center}
\includegraphics[width=4.2cm]{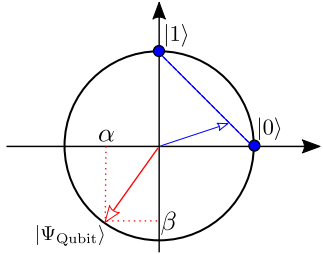}
\caption{The state of a qubit $\ket{\Psi_{\text{Qubit}}} = \alpha \ket{0} + \beta \ket{1}$ for real $\alpha$ and $\beta$ may be represented as a point on the unit circle (red arrow) as $|\alpha|^2+|\beta|^2=1$ for the probabilities to sum to unity. The possible states of a classical bit are equivalent to $\alpha=1$ or $\beta=1$ (blue circles). A qubit is different to a bit with classical probabilities to be 0 or 1: such states would be parameterised by two positive numbers that sum to one (blue arrow and blue dashed line).}
\label{Fig_qubit}
\end{center}
\end{figure}

\begin{figure}[h!]
\begin{center}
\subfigure[]{
\resizebox*{4.2cm}{!}{\includegraphics{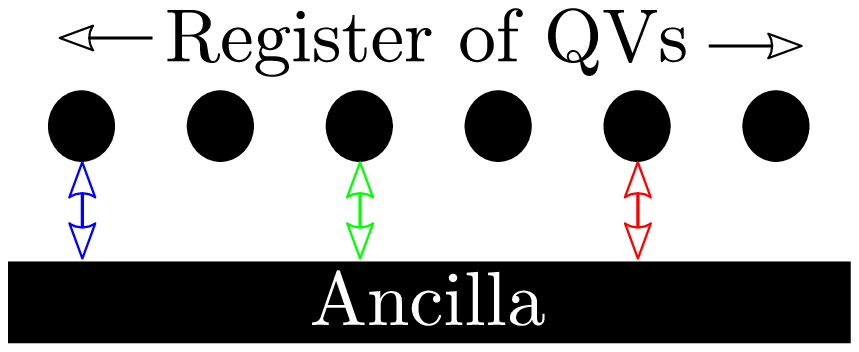}}}\hspace{12pt}
\subfigure[]{
\resizebox*{4.8cm}{!}{\includegraphics{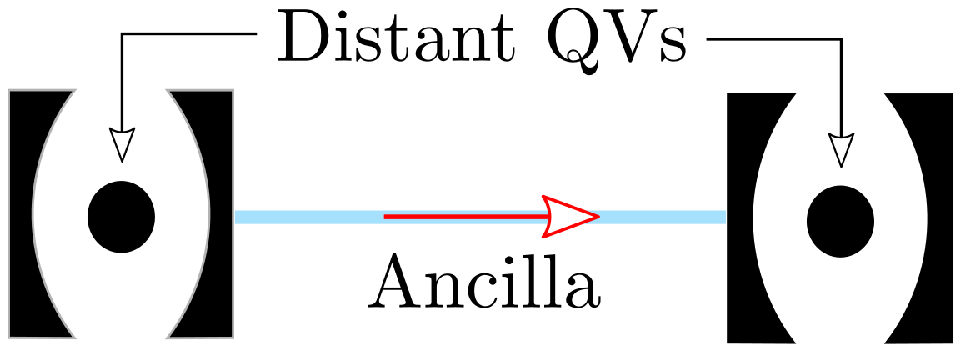}}}
\caption{An \emph{ancilla} or \emph{quantum bus} may be used to implement the interactions required for a quantum computation. (a) A quantum bus could be a distinct physical system which qubits may couple to in turn (e.g., flux qubits coupling to a superconducting resonator \cite{wang2009coupling,xue2012fast,spiller2006quantum}), denoted here by different colour arrows representing interations at different times. Alternatively, it might be a collective mode of several system qubits (e.g., vibrational modes of ions in an ion trap \cite{cirac1995quantum}). (b) Systems may interact via a `flying' quantum bus which is sent between distant systems, for example atoms interacting via photons.} \label{Fig_bus-schematic}
\end{center}
\end{figure}

The publication in 1994 of Shor's celebrated algorithm for efficient integer factoring using a quantum computer \cite{shor1994algorithms,shor1997polynomial} has sparked an explosion of interest in building such a device. For large numbers, this seemingly innocuous problem has so far proved impossible to solve in a reasonable length of time on a classical computer. Indeed, the assumed impracticality of this task is behind the security of widely used public-key cryptography methods, such as RSA encryption \cite{rivest1978method}. Although undermining current cryptography systems is little (if any!) motivation for undertaking the daunting tasking of actually building a quantum computer, since 1994 an expanding range of applications for such a device have been developed, including database searching \cite{grover1996fast}, machine-learning tasks \cite{schuld2015introduction}, and techniques for simulation of quantum systems \cite{brown2010using} as envisaged in the 1980s by Feynmann \cite{feynman1982simulating}. The degree to which quantum computers may enhance classical processing is a particularly interesting area of ongoing research: it is known that many tasks are not amenable to improved efficiency using a quantum computer \cite{aaronson2015read,aaronson2005guest,bennett1997strengths}.  Nonetheless, the known enhancements cover a wide range of important computational processes, and it is  likely that many more applications will become apparent if a fully-scalable quantum computer can be engineered.
\newline
\indent
What makes a quantum computer different from a classical computer? The overwhelming majority of modern classical computers are digital machines that encode information into a register of \emph{bits}, which may each take the values 0 or 1. A quantum computer instead consists of \emph{qubits} which may be in more general states which are neither definitely 0 nor 1 but in a wave-like \emph{superposition} of both. More specifically, a state of a qubit is described by a vector (using Dirac notation $\ket{.}$ for complex vectors)
 \begin{equation}
  \ket{\Psi_{\text{Qubit}}} = \alpha \ket{0 } + \beta \ket{1},
  \label{eq:qubit}
 \end{equation}
where $\ket{0}$ and $\ket{1}$ are orthonormal basis vectors and $\alpha$ and $\beta$ are complex numbers with $|\alpha|^2$ and $|\beta|^2$ the probabilities that the qubit is in the states $\ket{0}$ or $\ket{1}$ respectively when measured. The differences between a qubit and a classical bit are summarised in Fig.~\ref{Fig_qubit}. 
%
%
\newline
\indent
A classical computer stores data in a `register' of bits: for $N$ bits there are $2^N$ different possible states these bits may be in, so the register can represent up to $2^N$ different numbers.  At each step of the computation the register is in one of these states, e.g., one possible classical state is $R_{N\text{-bits}} = (010100\dots 0)$. In stark contrast, a quantum register may encode any superposition of these classical states simultaneously. Mathematically, the general state of a quantum register can be written as
\[
\ket{\Psi_{N\text{-qubits}}} = \sum_{x_q =0,1} \alpha_{x_1x_2...x_N} \ket{x_1x_2....x_N},
\]
where $x_1x_2...x_N$ are the individual bits (zero or one) and the sum runs over all permutations thereof. The coefficients assign a complex probability amplitude to each permutation and satisfy the usual normalisation condition $\sum |\alpha_{x_1,x_2,...,x_N}|^2 =1$. A quantum register can thus represent a superposition of all possible $N$-bit numbers at once.
\newline
\indent
A quantum computation consists of transformations between allowed quantum states
\[
\ket{\Psi_{N\text{-qubits}}}  \xrightarrow{}  \ket{\Psi'_{N\text{-qubits}}},
\]
and the transformations that do this are called \emph{unitary operators}. Because a quantum computer may be in a superposition of all possible bit strings at once, it may seem like it has access to an unreasonable level of parallelism. However, the output of a computation is given by measuring the qubit register at the end of the computation, which produces a single bit string with the probability $P(x_1x_2\dots x_N) =| \alpha_{x_1 x_2 \dots x_N}|^2$.  Hence, a quantum algorithm needs to intelligently make use of the allowed superpositions to enhance the probability of the desired result, illustrating the subtlety of quantum programming \cite{aaronson2015read,bacon2010recent}. To actually implement a given quantum computation described by some global $N$-qubit unitary, it must be decomposed into some physically available set of basic operations. Classical computations can be constructed from a small set of elementary logic operations that change the values of one or two bits at a time, e.g., negation (bit flip: $0\rightarrow 1$ and $1\rightarrow 0$) or AND (bit flip the second bit only if the first bit is a one). Similarly, there are small sets of elementary quantum gates that can generate all possible unitary operations, as will be discussed in Section~\ref{Sec_UQC}. 
\newline
\indent
Why is a quantum computer proving so hard to build? The central challenge in turning a quantum computer from an abstract curiosity into a real-world device is in sufficiently protecting the fragile quantum states from decoherence and errors during the computation.  Classical computers are designed with large error margins between their zero and one states, so mistakes are very rare and can be detected by simple data integrity `checksums' and parity checks. At the level of a few quantum particles, large error margins are not possible, and robust error correction is essential for a scalable quantum computer \cite{gottesman2010introduction,terhal2015quantum} . In addition to active error correction, quantum computer architectures must be designed to minimise errors through precise controls when manipulating the qubits, and to maximise coherence times by choosing naturally well-isolated quantum systems, such as nuclear spins \cite{zhong2015optically}, for quantum data storage. Fast quantum operations are also desirable, to maximise the number of gates that can be applied before decoherence builds up. These two requirements are incompatible: well-isolated implies difficult to interact with, and vice versa, so compromises must be made. One practical method of engineering interactions between well-isolated systems is by using a third system to mediate the interaction. Such mediating systems are often called a \emph{quantum bus} or an \emph{ancilla}, and can have different properties  that optimise them for interactions, in contrast with the well-isolated system qubits. Ancillas can be reset or discarded after a few gate operations, so they do not need to maintain coherence for the whole computation. By applying the more complex manipulations to the ancillas, then transferring the effects to the system qubits, the interactions with the system qubits can be simplified and minimised. These ideas are illustrated in the schematic diagram of Fig.~\ref{Fig_bus-schematic}.  
%
%
Ancillas are used in the gate designs of a wide range of promising physical systems being developed for qubit-based quantum computers.  The main topic of this review article is schemes of this sort, with a particular emphasis on the underlying unifying ideas. 
There are also radically different quantum computer architectures that do not use quantum gates, e.g., adiabatic quantum computing \cite{epstein2012adiabatic}, quantum annealing \cite{das2008colloquium,trummer2015multiple}, and a range of special purpose designs, e.g., for quantum simulations \cite{brown2010using}. These have their own advantages in the quest to overcome decoherence, but are outside of the scope of this review.
\newline
\indent
It is not essential to implement a quantum computer with 2-level systems (i.e., qubits). Indeed, most quantum systems have more than two dimensions and can thus in principle implement $d$-level logic or continuous information encoding.   
In Section~\ref{sec:hybrid}, qudits and quantum continuous variables are described and quantum computation with these systems is discussed. The advantages and disadvantages of using such systems are then summarised, and the ideas of hybrid quantum computing introduced.  This section may be read independently of the remainder of this paper as a basic introduction to the theory of quantum computing with these systems.  It covers the ideas necessary for understanding the ancilla-based gate methods that are the subject of Section~\ref{sec:ancilla}.  A brief overview of experimental implementations of such schemes is provided in Section~\ref{sec:expt}, and a summary of the key ideas concludes the paper in Section~\ref{sec:conclude}.

\section{Quantum computing in any dimension \label{sec:hybrid}}

In this section we review the terminology and mathematical tools for quantum computation using quantum systems of any dimension.  This allows the ideas of universal quantum computation to be explained in a unified manner and also highlights the advantages in using information encodings other than binary. 
%
%
%
\subsection{Qubits, qudits and quantum continuous variables \label{ssec:qudits}}
Classical digital computers need not be formulated with bits but may instead utilise any $d \in \mathbb{N}$ base logic, and indeed, according to D. E. Knuth (balanced\footnote{\emph{Balanced} ternary uses the values -1, 0 and 1 (as opposed to $0$, $1$ and $2$) and is hence naturally suited to representing negative numbers amongst other advantages.}) ternary logic is ``perhaps the prettiest number system of all'' \cite{knuth1968art}. A $d$-dimensional classical unit is called a \emph{dit}, and the equivalent quantum system is the $d$-dimensional \emph{qudit}. The general state for the $d=2$ qubit was given in Eq.~(\ref{eq:qubit}).  For a qudit of dimension $d$, the general state in Dirac notation can be written
\[ 
\ket{\Psi_{\text{Qudit}}} = \alpha_0 \ket{0} + \alpha_1 \ket{1} +\dots + \alpha_{d-1} \ket{d-1},
\]
where the $\ket{q}$ vectors are some `computational basis' (with $\expect{q | r}=\delta_{qr}$), and where $\alpha_q$ are complex numbers with $ |c_0|^2 + \dots + |c_{d-1}|^2 = 1$ (as $|c_q|^2$ is the probability to find the system in the state $\ket{q}$).  Rather than using Dirac notation, a qudit state may be given a perhaps more familiar and concrete representation in terms of an array of numbers, using the association
\[ \ket{0} = \begin{pmatrix} 1 \\ 0 \\ \vdots \\ 0 \end{pmatrix},\hspace{0.2cm}  \ket{1} = \begin{pmatrix} 0 \\ 1 \\ \vdots \\ 0 \end{pmatrix} ,\hspace{0.2cm}    \dots, \hspace{0.2cm} \ket{d-1} = \begin{pmatrix} 0 \\ 0 \\ \vdots \\ 1 \end{pmatrix},
\]
whence $\ket{\Psi_{\text{Qudit}}} = ( \alpha_0 , \alpha_1 , \dots,  \alpha_{d-1})^T$.

\begin{figure}[h!]
\begin{center}
\includegraphics[width=4cm]{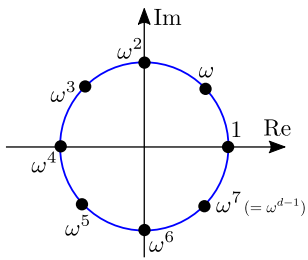}
\caption{The $d$ distinct $d^{th}$ roots of unity are integer powers of $\omega=e^{i 2\pi/d}$ and reside on the unit circle in the complex plane, illustrated here for $d=8$.} \label{dthrootofunit}
\end{center}
\end{figure}

The mathematics of finite-dimensional quantum theory (i.e., qudits) was initially developed by H. Weyl in the early decades of quantum mechanics \cite{weyl1950theory} and in the light of its relevance to quantum information has since been extensively investigated, with Vourdas \cite{vourdas2004quantum} providing an excellent technical review in a broader context. Only a basic understanding of qudit quantum mechanics is required to explain the fundamentals of quantum computation, which we now cover here. The $d$\textsuperscript{th} root of unity $\omega=e^{i \frac{2 \pi}{d}}$ plays an important role, for which $\omega^{0} +\omega^1+\omega^2+\dots+\omega^{d-1}=0$ as illustrated in Fig.~\ref{dthrootofunit}. 
%
%
Clearly, the value of $d$ determines the precise form of $\omega$ (as is also the case for all the objects introduced below), but everything discussed here holds true and is presented independently of its particular value. There are more advanced topics which depend on $d$ being prime (or a prime power), for example some phase space methods \cite{vourdas2004quantum,wootters1987wigner,gibbons2004discrete}, that are not needed here.
\newline
\indent
The basic operators in qudit quantum computation are the (generalised) \emph{Pauli operators} denoted $X$ and $Z$, which are the natural extension of two of the well-known \emph{qubit} Pauli operators $\sigma_x = \left( \begin{smallmatrix} 0&1\\ 1&0 \end{smallmatrix} \right)$ and $\sigma_z = \left( \begin{smallmatrix} 1&0\\ 0&-1 \end{smallmatrix} \right)$ respectively.  They may be defined (as can any operator) by their action on the computational basis states:
\[
X\ket{q} = \ket{q + 1}, \hspace{2cm} Z\ket{q} = \omega^{q}\ket{q},
\] 
where the addition is modulo $d$ as on a clock with $d$ hours, i.e., $(d-1)+1=0$. In terms of the more concrete column vector notation, they are the $d \times d$ matrices
\[  X = \begin{pmatrix}
  0 & 0 & \cdots & 0 & 1 \\
  1 & 0 & \cdots & 0 & 0 \\
    0 & 1 & \cdots & 0 & 0 \\
  \vdots  & \vdots  & \ddots & \vdots & \vdots  \\
  0 & 0 & \cdots & 1 & 0 \\
 \end{pmatrix}  , \hspace{1cm}
 Z = \begin{pmatrix}
 1 & 0 & 0 & \cdots  & 0 \\
 0 & \omega & 0 &\cdots  & 0 \\
 0 & 0 & \omega^2 &\cdots  & 0 \\
  \vdots   & \vdots & \vdots &\ddots & \vdots  \\
 0 & 0 & 0 & \cdots  & \omega^{d-1} \\
  \end{pmatrix} .
 \]
In fact, these operators pre-date quantum theory and were originally introduced by J. J. Sylvester in the $19$\textsuperscript{th} century \cite{sylvester2012collected} and in other contexts are called the `shift' and `clock' matrices respectively. The $X$ gate has a clear classic analogue which simply adds 1 modulo $d$ to a dit and is the natural extension of a bit flip ($0 \to 1$, $1\to 1+1=0$ modulo $2$).  In contrast, the $Z$ gate creates complex phase factors which do not have any obvious classical equivalent. 
\newline
\indent
These structures can be extended to continuous variables, which provide a quantum counterpart to classical  \emph{analog computers}.  Classical analog computers predate the invention of the digital computer, and are based on continuous degrees of freedom which take values on the real line $\mathbb{R}$ (in the ideal case).  The general theory of such machines was not developed until the work of Shannon in the mid-20\textsuperscript{th} century \cite{shannon1941mathematical}, which he based on J. Thompson's 19\textsuperscript{th} century mechanical differential analysers \cite{thomson1875mechanical}.  In the formative years of the digital computer, analog computers were still superior for a variety of tasks.  A \emph{quantum continuous variable} (QCV) \cite{kendon2010quantum,braunstein2005quantum} has a general state 
\[
\ket{\Psi_{\text{QCV}}} = \int_{-\infty}^{\infty} \text{d}q \, \psi(q) \ket{q},
\] 
where  $\ket{q}$ is some continuously parameterised state of the system, e.g., position in 1D, which again (in this context) is called the computational basis and $ \int_{-\infty}^{\infty} \text{d} q \,  | \psi(q)|^2  =1$ as $| \psi(q)|^2$ is the probability density for $q$. The $\psi(q)$ function is simply the quantum wavefunction familiar from elementary wave mechanics. The normalisation condition implies that the basis states themselves, i.e., the $\ket{q}$, are not physical as they are associated with a delta function $\psi(q') = \delta(q'-q)$ and the integral of $|\delta(q'-q)|^2$ is divergent\footnote{Any wavefunction for which its modulus squared integrates to a constant can be normalised, these are called the \emph{square-integrable} functions and they are the physically allowed states.}. However, they can be approximated, for example by a Gaussian wave function centred on $q$ with a narrow peak, known as a \emph{squeezed state} \cite{braunstein2005quantum}. The basic operators in QCV quantum computation are again known as the Pauli operators and are given by
\[
X(q') \ket{q} = \ket{q+q'} ,\hspace{1cm} Z(q') \ket{q} = e^{i q q' }\ket{q},
\]
where as before, the $X$ gate has the natural classical analogue of addition. In certain contexts (especially quantum optics), these operators are often replaced by the entirely equivalent \emph{displacement operators}, with the difference simply one of convention.\footnote{The displacement operator is given by $\mathcal{D}(\alpha) \equiv \sqrt{\omega^{qq'}} Z^{q'} X^{q}$ with $\alpha=(q+iq')/\sqrt{2}$ a complex parameter. It is also possible to define displacement operators for a qudit in the same manner \cite{vourdas2004quantum}.}  The Pauli $X$ and $Z$ gates are more convenient for quantum computing because they are the natural notation for \emph{quantum error correction}  \cite{gottesman2010introduction,raussendorf2012key,terhal2015quantum}.  While essential for practical quantum computers to mitigate the effects of decoherence on fragile quantum states, quantum error correction is beyond the scope of this review.
\newline
\indent
It is clear that the basic structures for qubits, qudits and QCVs are all equivalent. This can be made particularly evident in the following way: for a qudit let $ X(q)  \equiv X^q$, $ Z(q) \equiv Z^q$ with integer $q$, and for a QCV define the dimensionality constant $d$ to be $d=2\pi$. Then in all cases the Pauli operators obey
\begin{equation}
X(q')\ket{q} = \ket{q + q'}, \hspace{2cm} Z(q')\ket{q} = \omega^{qq'}\ket{q},
\label{XZactioncomp}
\end{equation}
for $q$ and $q'$ restricted to integers for qudits and any real number for QCVs, as will be implicitly assumed from now on. Note that here we still have that $\omega^{2\pi i /d}$. These ideas allow us to consider quantum computation in a largely dimension-independent fashion, which will be particularly useful when describing ancilla-based models in Section~\ref{sec:ancilla}. The Pauli operators play a central role in quantum computation due to their many useful properties. One property that will be used later is that they commute up to a phase, specifically
 \begin{equation} 
 Z(q) X(q')= \omega^{qq'} X(q') Z(q).
 \label{eq:paulicommute}
 \end{equation}
Confirming this is the case for qubits ($\omega=-1$, $q,q'$ are integers) via matrix multiplication of the ordinary $2\times 2$ qubit Pauli gates is easily done, and for qudits and QCVs is still straightforward.  This commutation relation is essential in the definition of the \emph{Pauli group} and hence the \emph{Clifford group} \cite{gottesman1999heisenberg,bartlett2002efficient,gottesman1999fault} which play a key role in, for example, the theory of error correction and fault tolerance \cite{gottesman2010introduction,terhal2015quantum}, measurement-based quantum computation \cite{browne2006one,zhou2003quantum,menicucci2006universal}, and in understanding the fine line between universal quantum computation and efficiently classically simulatable quantum computation \cite{gottesman1999heisenberg,bartlett2002efficient,van2013efficient,gottesman1999fault}.

\subsection{Translations in phase space}
\begin{figure}[h!]
\begin{center}
\includegraphics[width=5.0cm]{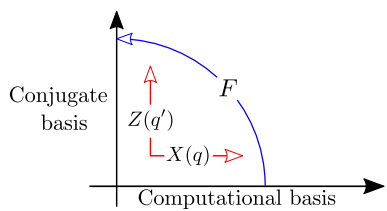}
\caption{Powers of the Pauli operators may be represented as translations in phase space along computational basis and conjugate basis axes. The Fourier transform is a $\pi/2$ rotation in phase space. For a QCV the background phase space is the plane of real numbers and for a qudit it is a $d \times d$ discrete (and periodic) lattice. There is a wide range of rigorous and powerful phase space methods, e.g., quasi-probability distribution functions \cite{silberhorn2007detecting,leonhardt2005measuring}. However, here, phase space will be used only as an intuitive schematic aid.} \label{Fig_phasespace}
\end{center}
\end{figure}

In order to provide an intuitive picture for the action of the Pauli operators, and as an essential ingredient in the ancilla-based models that are the main subject of this review, it will be helpful to introduce the Fourier transform gate $F$. As its name suggests, it is simply the unitary representation of discrete and continuous Fourier transforms for qudits and QCVs respectively. Hence, for a qudit it is given by  
\[ 
F \ket{q}= \frac{1}{\sqrt{d}}\sum_{q'=0}^{d-1}\omega^{qq'}\ket{q'},
\] 
and for a QCV the sum is replaced by an integral, with
\[
F\ket{q} = \frac{1}{\sqrt{2\pi}} \int_{-\infty}^{\infty} dq' e^{ i q q'} \ket{q'}.
\]
Note that for qubits, $F= \left( \begin{smallmatrix} 1&1\\ 1&-1 \end{smallmatrix} \right)/\sqrt{2}$, which is more commonly known as the Hadamard gate and often denoted $H$.
For a QCV, it is a particularly natural operator in any system governed by the quantum harmonic oscillator Hamiltonian, e.g., a micro-mechanical resonator \cite{poot2012mechanical} or a single light mode \cite{gerry2005introductory,radmore1997methods}. This operator is ubiquitous in quantum circuits, and its multi-system generalisation is a key ingredient in many quantum algorithms, e.g., Shor's algorithm \cite{shor1994algorithms,shor1997polynomial}. By applying the Fourier transform to the computational basis the \emph{conjugate basis} states are obtained, denoted $ \ket{+_q} \equiv F\ket{q}$.  They are equal superpositions of all possible computational basis states, e.g., for a qudit $\ket{+_1}$ is
\[ 
\ket{+_1} =\frac{1}{\sqrt{d}} \left( \ket{0} + \omega \ket{1} + \dots + \omega^{d-1} \ket{d-1} \right).
\]
They are states of maximal uncertainty in terms of the outcomes of computational basis measurements (and vice versa) which is a property inherited directly from their relationship to a Fourier transform, and hence important in areas such as measurement-driven gates (as will be seen later) and teleportation \cite{bennett93teleporting}. In the conjugate basis, the roles of the Pauli operators are reversed, more precisely 
\begin{equation} 
Z(q')\ket{+_q} = \ket{+_{q+q'}}, \hspace{2cm} X(q')\ket{q} = \omega^{-qq'} \ket{+_q}.
\label{Eq-conjP}
\end{equation}
The actions of the Pauli and Fourier transform operators can be intuitively summarised in phase space as shown in Fig.~\ref{Fig_phasespace}. 
%
%

\subsection{Interactions between quantum systems \label{Sec_interact}}

So far, all the operations described have applied to one quantum system - a qubit, qudit or QCV system.  For quantum computing, there is one further crucial ingredient: interactions between two quantum systems.  The quantum gate often considered is the $\textsc{sum}$ gate (called \textsc{cnot} for qubits) given by
\begin{equation}  
\textsc{sum}  \ket{m}\ket{n}  = \ket{m}\ket{m+n} ,
\label{eq:sum}
\end{equation}
as adding the value of two quantum variables is naturally a useful computational resource. It is clear that $\textsc{sum}$ may entangle the two quantum systems as, for example,
\[
\frac{1}{\sqrt{d}}(\ket{0} + \ket{1} + \ket{ 2} \dots ) \ket{0}  \xrightarrow{\textsc{sum}}  \frac{1}{\sqrt{d}}(\ket{00} + \ket{11} + \ket{ 22} \dots ),
\]
and for qubits this is $\frac{1}{\sqrt{2}}(\ket{00}+\ket{11})$ which is one of the famous Bell-states, at the heart of many quantum information protocols (e.g., teleportation \cite{bennett93teleporting} and cryptography \cite{hughes1995quantum}) and the modern formulation of the hotly-debated Einstein-Podolski-Rosen paradox \cite{einstein1935can}.  
\newline
\indent
The $\textsc{sum}$ gate is a particular case of an important class of gates that are ubiquitous in quantum computation: the controlled-$u$ gates, denoted $Cu$, with the action 
\begin{equation}
  \ket{m}\ket{n}  \xrightarrow{Cu} \ket{m} u^m\ket{n},
\label{eq:Cu}
\end{equation}
for some unitary $u$. Hence, $\textsc{sum}=CX(1)$. Controlled-gates can be defined on hybrid variables (e.g., a `control' qubit and a `target' QCV, with $u$ a gate that acts on a QCV), and such hybrid gates will be used throughout Section~\ref{sec:ancilla}. Having reviewed the basics operations for qubits, qudits and QCVs we now turn our attention to quantum computation.

\subsection{Universal quantum computation \label{Sec_UQC}}
\begin{figure}[h!]
\begin{center}
\includegraphics[width=7.5cm]{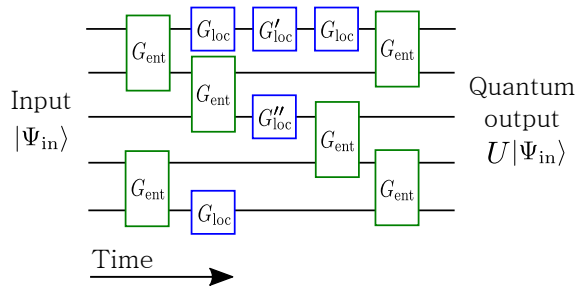}
\caption{A circuit diagram may be used to represent a gate sequence, where a wire represents each subsystem, time flows from left to right and boxes represent gates. Any quantum computation may be implemented on a register using repeated application of a single two-body entangling gate (green boxes) and local gates (blue boxes) alone.} \label{Fig_UQC}
\end{center}
\end{figure}

A \emph{universal quantum computer} consisting of qudits is a device which can implement \emph{any} global unitary operation $U$ on a prepared quantum input state of the qudits $\ket{\Psi_{\text{input}}}$, and measure the result,
\[
\ket{\Psi_{\text{input}}} \xrightarrow{U} \ket{\Psi_{\text{output}}}\xrightarrow{\text{measure}} \text{Classical output}.
\]
The situation for QCVs is a little more subtle, and instead of needing to implement any unitary only a physically reasonable subset is considered. This technicality is ignored for now and is explained in more detail below. 
\newline
\indent
It turns out that, like classical digital and analog computation, any quantum computation can be implemented by a sequence of elementary operations. Moreover, the only basic gates that are required for universal quantum computation are \emph{any} two-body entangling gate, combined with a choice of \emph{local} gates sufficient to implement, or approximate, any other local gate \cite{brylinski2002universal,Lloyd1999quantum} (a local gate is one that acts on a single subsystem, e.g., $F$). This idea is summarised in the \emph{circuit diagram} of Fig.~\ref{Fig_UQC}. 
%
%
The gates so far introduced ($\textsc{sum},X(q),Z(q),F$) are \emph{not} a sufficient set of gates to implement any quantum computation (implied by the Gottesman-Knill theorem and its extensions \cite{gottesman1999heisenberg,bartlett2002efficient,hostens2005stabilizer,van2013efficient,gottesman1999fault}).  The simplest additional gate for qubits is the $T$ gate, which adds a phase of $\pi/8$ to the $\ket{1}$ component of a qubit, and more generally gates which implement appropriate phases can elevate this set to universality in all dimensions \cite{Lloyd1999quantum,Proctor2015ancilla,howard2012qudit}. However, the choice of elementary gates is often determined by the most natural operations in the particular quantum system, see Section~\ref{sec:expt}, and where it is possible to implement a larger range of local gates this will enable more efficient quantum programs. 
\newline
\indent
As already noted, for QCVs the computational basis contains a continuum of basis states. Hence, an arbitrary unitary transformation is defined by a continuum of parameters for QCVs, and cannot therefore be decomposed into a finite sequence of computational gates. An alternative definition of universal quantum computation for QCVs was give by Lloyd and Braunstein \cite{Lloyd1999quantum} in 1999.  It works by constructing a physically reasonable set of unitaries from the bottom up: The Pauli operators for a QCV may be related to the Hermitian position and momentum operators $\hat{x}$ and $\hat{p}$ which obey the \emph{canonical commutation relation} $[\hat{x}, \hat{p}] = i$, via $X(q)=e^{ -iq \hat{p}}$ and $Z(q) = e^{iq\hat{x}} $.  The computational and conjugate basis are the eigenstates of $\hat{x}$ and $\hat{p}$ respectively.  Using these operators, a good definition of a universal QCV computer is one which by composing its basic gates can implement any unitary of the form
\[U=e^{i \text{poly} (\hat{x}_1 , \hat{p}_1, \dots ,\hat{x}_N , \hat{p}_N )},
\]
where $\text{poly} (\hat{x}_1 , \hat{p}_1, \dots ,\hat{x}_N , \hat{p}_N )$ is any real polynomial in the $\hat{x}$ and $\hat{p}$ operators of each of the $N$ QCVs. For example, a possible polynomial for two QVCs is $a \hat{x}_1^3 + b \hat{p}_1^2 \hat{p}_2 + c \hat{p}^7_2$ for real $a,b$ and $c$. This set of unitaries is sufficient to implement useful quantum computations \cite{Lloyd1999quantum}.
\newline
\indent
As we've now seen, the mathematics of quantum computation is pretty much the same whatever the dimension of the underlying quantum system.  But qubits are clearly simplest, so there must be good reasons to consider using higher dimensional systems.  One good physical motivation is that many quantum systems naturally allow for a qudit or QCV encoding, for example, atoms and ions have many electronic energy levels; coherent states of light or other electromagnetic radiation are the archetypal QCVs, and are some of the most straightforward experimental systems to prepare and manipulate: discussions of relevant proposals and experimental progress in this area are deferred to Section~\ref{sec:expt}.  And there are also more abstract advantages. For qudits, there is the potential for a $\log_2(d)$ reduction in the number of gates and subsystems required for a computation, although this is countered by the increased complexity of each gate and any advantages would depend on the details of a given set-up \cite{Muthukrishnan2000Multivalued,stroud2002quantum}. Moreover, a further advantage is that qudit algorithms have been shown to exhibit increased robustness and success probability \cite{parasa2011quantum,zilic2007scaling,parasa2012quantum} and particularly striking recent results show that qudit quantum error correcting codes possess remarkable improvements with increased qudit dimension \cite{watson2015qudit,campbell2014enhanced,anwar2014fast,campbell2012magic,andrist2015error,duclos2013kitaev}. Turning to QCV, some problems are clearly most naturally encoded using continuous parameters. However, QCVs are potentially most suited to a type of \emph{hybrid} quantum computation which uses different types of encoding simultaneously, e.g. qubits combined with QCVs. Such a device could have the advantages of discrete and continuous variables simultaneously and this has been used to construct simpler algorithms for finding eigensystems  \cite{lloyd2003hybrid,travaglione2001generation}.  
A natural choice for a quantum bus is a QCV system, and there are diverse examples that have been proposed and implemented \cite{cirac1995quantum,munro2005weak,wang2009coupling,xue2012fast,spiller2006quantum,armour2002entanglement}. 
The next section will explore the options for hybrid- and ancilla-based quantum computing in more detail.

\section{Ancilla-based quantum computation \label{sec:ancilla}}
\begin{figure}[h!]
\begin{center}
\includegraphics[width=10cm]{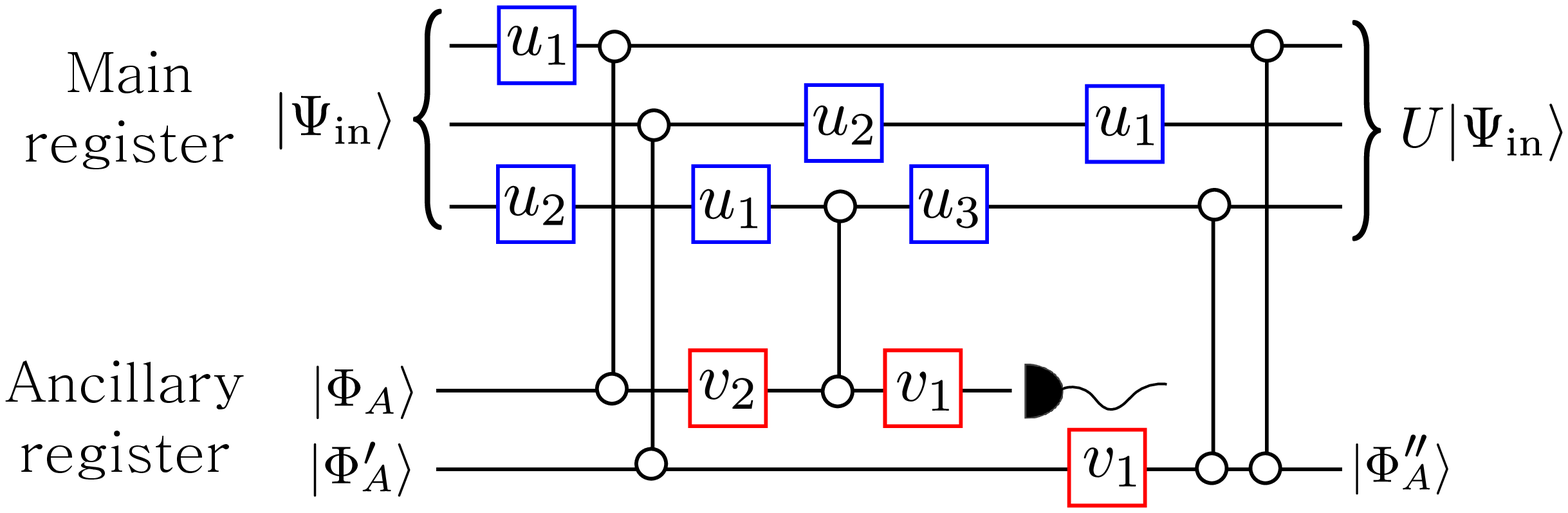}
\caption{A circuit diagram illustrating the ideas of universal quantum computation via ancilla-mediated gates. The elements of a main computational register and ancillary register interact via some ancilla-register interaction gates (black connecting lines). Further computational resources that may be required are local gates on the main register (blue boxes), local gates on the ancilla (red boxes), ancilla measurements (black measuring devices) and ancilla state preparation (controlling the $\ket{\Phi_{A}}$ states).} 
\label{Fig_ancilla-based-circuits}
\end{center}
\end{figure}

Implementing quantum computation on a register requires both isolating individual subsystems to minimise decoherence, and applying precisely controlled interactions between subsystems to implement the computation. The tension between these demands is one motivation for sidestepping direct interactions and instead mediating interaction gates via an ancillary system. This allows the register to be specifically tailored for long coherence times and interactions are only required with some physically distinct ancillary systems that naturally interact with the elements of the main register and which exhibit complementary properties, such as easy manipulation. Another big advantage is that the ancilla can mediate interactions between register subsystems that are not next to each other, thus avoiding the need to move subsystems around, or swap their states onto adjacent subsystems for direct interactions. Clearly, the register and ancillas need not be of the same type. Hence, this is a natural setting for hybrid quantum computation. Implementing universal quantum computation via ancillas may be summarised by the schematic circuit diagram of Fig.~\ref{Fig_ancilla-based-circuits}. 
%
%
\newline
\indent
There are essentially two basic techniques that are used to implement gates via ancillas. The first is \emph{delocalising} the quantum information in a register system across the system and an ancilla. To illustrate this for qubits: A general state of a register qubit is given by $\ket{\Psi} = \alpha\ket{0} +\beta\ket{1}$, and by preparing an ancilla in the state $\ket{0}$ and interacting via $\textsc{cnot}$ (the qubit $\textsc{sum}$ gate) then
\[ 
 (\alpha \ket{0} +\beta \ket{1}) \ket{0} \xrightarrow{\textsc{cnot}} \alpha \ket{0}\ket{0} +\beta \ket{1} \ket{1}  .
 \]
 The information encoded in the values of $\alpha$ and $\beta$ then resides non-locally in both qubits and manipulations of the ancilla will affect the state of the \emph{logical} qubit ($\alpha\ket{0} + \beta\ket{1}$) and can be used to entangle it with further register qubits. To complete a gate of this type, the logical qubit must be relocalised into the register (here this can be achieved by applying a second $\textsc{cnot}$ gate). This technique is the basis of all of the gate methods reviewed in Sections~\ref{Sec_Geo-phase-gates} to \ref{Sec_measured-gates} and much of Section~\ref{sec:minimal}. An alternative to delocalising the quantum information stored in a register systems is to transfer it completely into the ancillary system, i.e.,
 \begin{equation}\label{eq:swap}
 \ket{\Psi} \ket{\Phi_A} \xrightarrow{\textsc{swap}}   \ket{\Phi_A}\ket{\Psi},
 \end{equation}
defining the $\textsc{swap}$ gate. Obviously, manipulations of the ancilla will then transform the logical $\ket{\Psi}$ state, which must then be swapped back into the register to complete the gate. The $\textsc{swap}$ gate cannot create entanglement, and hence cannot be used as the only two-body gate in quantum computation. Nevertheless, this technique may be utilised in a slightly more subtle fashion for resource-efficient ancilla gates, as will be seen in the latter part of Section~\ref{sec:minimal}.

%
%
\subsection{Geometric phase gates \label{Sec_Geo-phase-gates}}
\begin{figure}[h!]
\begin{center}
\includegraphics[width=7.7cm]{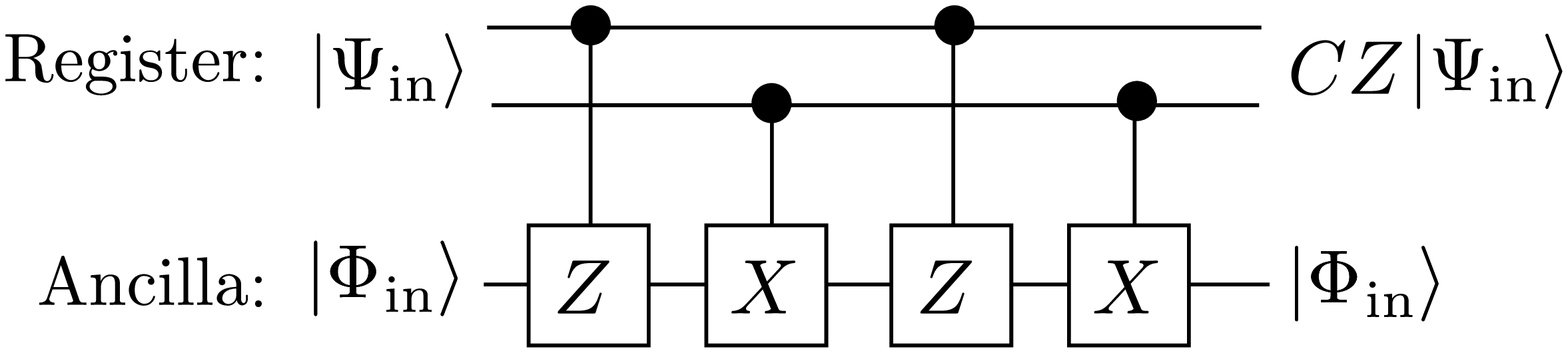}
\caption{Two register qubits may be entangled via $CZ$ and $CX$ (\textsc{cnot}) gates acting on an ancilla qubit.} \label{Fig_qubit-phase-gate}
\end{center}
\end{figure}

\begin{figure}[h!]
\begin{center}
\includegraphics[width=5.3cm]{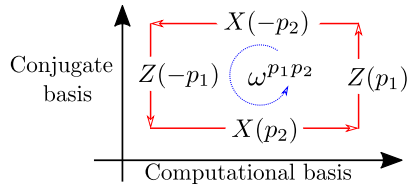}
\caption{A closed loop of Pauli operators creates an area-dependent geometric phase} \label{Fig_geophase}
\end{center}
\end{figure}

\begin{figure}[h!]
\begin{center}
\includegraphics[width=7cm]{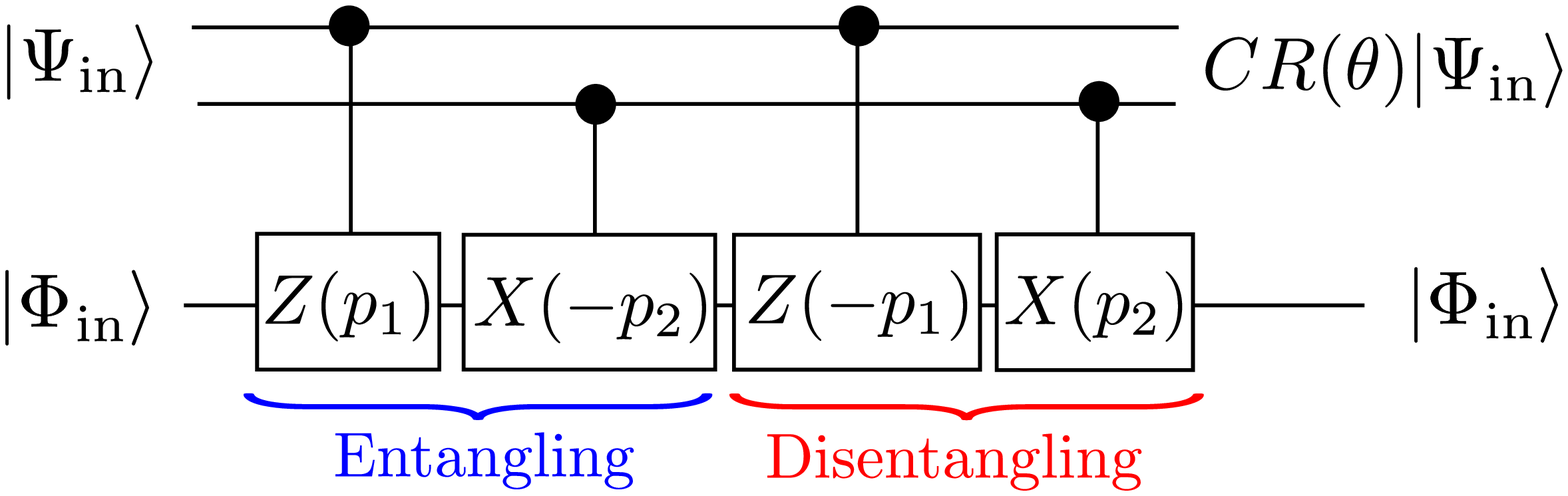}
\caption{Register systems may be entangled via an ancilla using register-controlled Pauli gates. This translates the ancilla around a closed phase space loop with the area dependent on the state of the register systems, and hence effects an entangling gate on the register. It will be seen later that the `disentangling stage' may be replaced by a measurement of the ancilla.} \label{Fig_phase-gate}
\end{center}
\end{figure}

Register-controlled Pauli gates on ancillas may be used to construct a particularly useful class of ancilla gates: \emph{geometric phase gates}.  We will first illustrate this using qubits for the register and ancillas. Consider the gate sequence of Fig.~\ref{Fig_qubit-phase-gate}. What is its action on the ancilla? Well, if the two register qubits are in the states $\ket{q_1}$ and $\ket{q_2}$ respectively, then using the definition of a controlled gate (see Eq.~(\ref{eq:Cu})) the operator applied to the ancilla is
\[
 \begin{pmatrix} 0&1\\ 1&0 \end{pmatrix}^{q_2} \begin{pmatrix} 1&0\\ 0&-1 \end{pmatrix}^{q_1}  \begin{pmatrix} 0&1\\ 1&0 \end{pmatrix}^{q_2} \begin{pmatrix} 1&0\\ 0&-1 \end{pmatrix}^{q_1} = (-1)^{q_1q_2}\begin{pmatrix} 1&0\\ 0&1 \end{pmatrix}.
\]  
Hence, it has no net effect (i.e., an identity) on the ancilla, but regardless of the ancilla state adds a $-1$ phase if $q_1=q_2=1$. This is exactly the action of the two-qubit entangling gate $CZ$ on the two register qubits. Hence, by interacting each qubit with an ancilla twice, an entangling gate between the two register qubits has been mediated - this is precisely the idea of ancilla-based gates. This gate then facilitates universal quantum computation when augmented by a sufficient set of local gates.
%
%
\newline
\indent
This same basic scheme can mediate gates with register and ancilla of any (and possibly different) variable types using register-controlled Pauli gates on the ancilla: the only restriction is that if the register consists of QCVs the ancillas must also be QCVs as will be implicitely assumed in the following. This gate was originally proposed by Milburn \cite{milburn1999simulating} in the context of a QCV bus and a qubit register, and this idea has been further developed in \cite{van2008hybrid,proctor2014quantum,milburn2000ion,louis2008loss,spiller2006quantum,wang2002simulation,louis2007efficiencies,brown2011ancilla,horsman2011reduce,khosla2013quantum}. In general, the Pauli gates commute up to a phase as was seen in Eq.~(\ref{eq:paulicommute}), and that relation clearly implies that
\begin{equation}\label{eq:gphase}
X(p_2) Z(-p_1)X(-p_2)Z(p_1) = \omega^{p_1p_2}\identity.
\end{equation}
This may be understood pictorially in phase space as a closed loop creating an area-dependent phase as shown in Fig.~\ref{Fig_geophase}.
%
%
Global phases have no physical consequence in quantum mechanics, however, by repeating the qubit recipe above, this can be turned into a \emph{controlled} geometric phase. This is achieved by acting these gates on an ancilla with the $Z(\cdot)$ Pauli gates controlled by one register system and $X(\cdot)$ Pauli gates controlled by a second register system. 
This implements an entangling gate between the two register systems as shown in the circuit diagram of Fig.~\ref{Fig_phase-gate}. Specifically, for the register systems in the states $\ket{q_1}$ and $\ket{q_2}$ respectively, the action on the ancilla is
\[
X(q_2p_2) Z(-q_1p_1)X(-q_2p_2)Z(q_1p_1) =  \omega^{p_1p_2 q_1q_2} \identity,
\]
which is a phase controlled by the state of the two register systems. Such a gate is in general called a \emph{controlled phase gate}, denoted $CR(\theta)$ where $R(\theta)\ket{q}=e^{i \theta q} \ket{q}$ is parameterised by the phase angle $\theta$. The exact gate enacted on the register here has a parameter $\theta$ fixed by the gate variables $p_1$ and $p_2$ and the variable type of the ancilla (in general, $\theta=2 \pi p_1p_2/d$). For clarity, we consider the three cases individually: 
\begin{itemize}
\item Qubit ancilla: $\theta=p_1p_2 \pi $ with $p_1$ and $p_2$ integers.  
\item Qudit ancilla (with dimension $d$): $\theta= 2 p_1p_2\pi/d$ with $p_1$ and $p_2$ integers. 
\item QCV ancilla: $\theta=p_1p_2$ with $p_1$ and $p_2$ taking any real values. 
\end{itemize}
In all cases, this is sufficient for universal quantum computation when augmented with local controls. However, the ancilla type restricts what controlled phase gates may be applied to the register, with less restriction as the dimension of the ancilla increases, e.g, a qubit ancilla can only implement $\pm 1$ phases. In the QCV ancilla case, the gate parameters may be chosen to implement a $CZ$ gate for a register of any variable type as there is total freedom to pick the value of $\theta$. In all cases, as the ancilla is left unchanged by the gate, it may be either reused, discarded or reset to remove any residual entanglement from imperfect operation. This analysis considers only the ideal case of perfect implementation, and in reality decoherence of the ancilla during gate implementation will be unavoidable in experimental settings, resulting in decoherence in the systems coupled to the ancilla and imperfect gate implementation.  However it has been shown by Louis \emph{et al.} \cite{louis2008loss} that for a photonic ancilla this gate method is sufficiently robust for low to medium decoherence.
%
%

\subsection{Efficient gate compositions}
\begin{figure}[h!]
\begin{center}
\includegraphics[width=11cm]{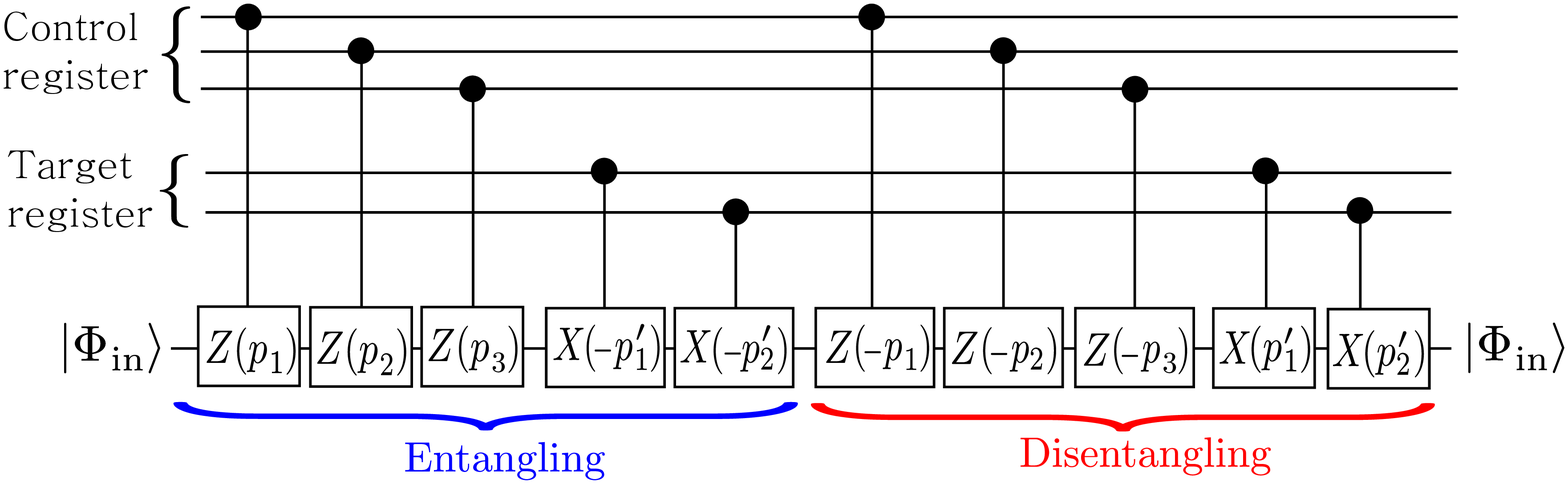}
\caption{Illustration of how gate sequences can be compressed for suitable sequences of gates.  This enacts a controlled rotation between each control and target pair of register systems.} \label{Fig_efficiencies}
\end{center}
\end{figure}

\begin{figure}[h!]
\begin{center}
\includegraphics[width=8.0cm]{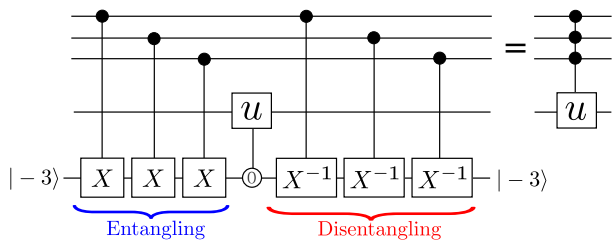}
\caption{The $N$-qubit generalised Toffoli gate applies some given unitary $u$ to a target qubit only if all $n$ control qubits are in state $\ket{1}$. This may be implemented using only $2N+1$ interactions with an ancillary qudit with dimension $d >N$ using the sequence shown here for $N=3$. The qudit controlled gate implements $u$ on the register qubit when the qudit is the state $\ket{0}$.} \label{Fig_ancilla-Toffoli-gate}
\end{center}
\end{figure}

The geometric phase gate described above is sufficient for universal quantum computation on the register (assuming the addition of local controls), and hence any gate sequence can be implemented by repeated application of such gates. Moreover, the gate can be adapted 
to  implement some common gate sequences in a more efficient fashion. It is essentially trivial that
\begin{equation}
Z(q_N) \dots Z(q_2) Z(q_1)  =   Z(q_N+\dots+q_2+q_1),
\label{Eq:pauli-comb}
 \end{equation}
and similarly for $X(q)$ gates. Hence, consider the gate sequence in Fig.~\ref{Fig_efficiencies} in which many variables, separated into `control' and `target' registers of $N$ and $M$ variables respectively, interact in turn with an ancilla. If the $j$\textsuperscript{th} variable in the control and target registers are in the state $\ket{c_j}$ and $\ket{t_j}$ respectively, then using Eq.~(\ref{Eq:pauli-comb}) and then applying Eq.~(\ref{eq:gphase}) it may be confirmed that this sequence maps the ancilla to
\[
\ket{\Phi_{A} } \to \omega^{(p_1 c_1+\dots +p_{N} c_{N} )(p'_1 t_1+\dots +p'_{M} t_{M} )} \ket{\Phi_{A}}.
\]
Expanding the brackets shows that this is $N \times M$ controlled rotation gates: one between each of the systems in the control register and each of those in the target register. Hence this implements a number of gates that is quadratic in the number of target and control systems ($N \times M$) using only a linear number of operations ($2N+2M$), creating a substantial saving in the number of two-body interactions over a gate-by-gate method. However, the rotation parameters in the gates implemented cannot all be independent \cite{brown2011ancilla} -- there are only $N+M$ parameters in the ancilla-mediated gate sequence. Furthermore, the range of gates that can be implemented depends on the dimensionality of the ancilla (e.g., if it is a qubit all gates are either the identity or $CZ$ as $\omega=-1$)  \cite{proctor2014quantum}, with increasing freedom with increasing dimension. This illustrates the subtle nature of any advantages gained from using a higher-dimensional ancilla. In the context of a qubit register and QCV ancilla, this principle has been applied to design more intricate sequences of operations for increasing efficiencies in quantum simulation \cite{brown2011ancilla} and for making \emph{cluster states} \cite{brown2012layer,horsman2011reduce,louis2007efficiencies} with comparisons with what can be achieved using a qudit ancilla given in \cite{proctor2014quantum}. Equivalent ideas directly carry over to a register of qudits although this has yet to be investigated in detail.
%
\newline
\indent
Similar methods can be used for a highly efficient implementation of a \emph{generalised Toffoli} gate on a register of \emph{qubits} using a \emph{qudit} bus \cite{ionicioiu2009generalized}. The generalised Toffoli gate has $N$ control qubits and a single target qubit, and implements the unitary $u$ on the target only if \emph{all} the control qubits are in the state $\ket{1}$, i.e. it is the map
\[  
 \ket{c_1 \dots c_N}\ket{t} \xrightarrow{\textsc{toffoli}} \ket{c_1 \dots c_N}  u^{c_1 \times\dots \times c_N} \ket{t}.
 \]
The Toffoli gate plays an important role in quantum computation, for example, it appears in many error correcting codes \cite{gottesman1997stabilizer} and is a natural component in a variety of quantum algorithms \cite{nielsen2010quantum}.  Hence efficient decompositions are of interest. The importance of the Toffoli gate is linked to the fact that the ordinary Toffoli gate ($N=2$, $u=X$) is a valid classical 3-bit gate and alone is universal for classical reversible computation \cite{nielsen2010quantum}. The addition of only the Hadamard gate, or indeed \emph{any} basis changing gate \cite{shi2002both}, is enough to make this universal classical set become universal for quantum computation.  A very simple technique may be employed to efficiently implement this gate via a qudit ancilla: the ancilla is used to count the number of qubits in the $\ket{1}$ state. To write this value into the ancilla it is initialised to $\ket{-N}$ (i.e., to $d-N\equiv -N$ \textsc{mod} $d$) and the `entangling' stage of Fig.~\ref{Fig_ancilla-Toffoli-gate} is applied, mapping
\[   \ket{c_1 \dots c_N}\ket{t} \ket{-N} \xrightarrow{} \ket{c_1 \dots c_N} \ket{t}  \ket{c_1+\dots+c_N-N} .  \]
Hence, the bus is in the state $\ket{0}$ when $c_1+\dots+c_N=N$, i.e., when all of the qubits are in the $\ket{1}$ state, as long as $d>N$ (otherwise the modulo $d$ nature of the addition will come into play). Therefore, if a gate is implemented on the target subsystem controlled on whether the ancilla is in the state $\ket{0}$ this implements the required $u^{c_1\times \dots \times c_N}$ gate on the target. By inverting the counting stage, the gate is completed. This requires $2N+1$ gates, which is more efficient than any known scheme using qubits alone \cite{saeedi2013linear}. A variation on this method has been demonstrated experimentally with photonic qubits \cite{lanyon2009simplifying}. These ideas are not dependent on the register consisting of qubits and may be adapted to many-qudit gates, although as far as we are aware this has not been investigated. This idea also applies in principle for a QCV bus. However, in this case the scheme requires a gate which only implements $u$ on the register if the bus is exactly in the state $\ket{0.000\dots}$ and such a gate is unphysical.  Suitable approximations may be valid and useful, but have not so far been investigated.
%
%
\newline
\indent
Correlations between errors on different register systems can cause problems for quantum error-correction \cite{terhal2015quantum} and errors of this sort will be created if many systems are coupled simultaneously to a bus. There will then be a trade-off between reducing gate counts and introducing these problematic errors, and this optimisation has been considered by Horsman \emph{et al.} \cite{horsman2011reduce} confirming that gate-count reductions of this sort may indeed prove useful in practice.
\newline
\indent
Ancilla gates use (in general) hybrid variables and hence are already hybrid quantum computation in one sense. However, the focus has been entirely on implementing gates on the register and has not considered whether some computation may also be implemented explicitly in the ancilla. Since many ancillas will be needed to implement gates in parallel: the computation will employ a main computation register and an ancillary register, as was shown in Fig.~\ref{Fig_ancilla-based-circuits}. Interestingly, the ancilla-register interaction gates considered so far (hybrid-variable controlled Paulis) along with local controls also allow universal quantum computation on both the main and the ancilla registers, i.e., may be used for truly hybrid computation \cite{lloyd2003hybrid}.

\subsection{Entangling gates using ancilla measurements \label{Sec_measured-gates}}

\begin{figure}[h!]
\begin{center}
\includegraphics[width=6cm]{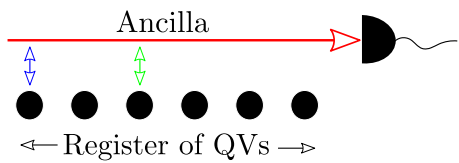}
\caption{A schematic of two register systems interacting sequentially with a bus followed by a measurement of the bus. An entangling gate may be implemented on the register in this fashion. } \label{Fig_schematic_busm}
\end{center}
\end{figure}

The geometric phase gate requires four ancilla-register interactions: two with each register system. This is because both register systems entangle with (or delocalise into) the ancilla; to disentangle them requires two further interactions. This may be particularly inconvenient if, for example, the register systems are distant and the bus is a light pulse sent between them. To avoid repeated interactions with the same ancilla requires going beyond unitary gates and using a measurement of the ancilla, see figure \ref{Fig_schematic_busm}. 
%
%

Consider again register and ancilla qubits and the gate sequence of Fig.~\ref{Fig_qubit-phase-gate}. If the ancilla is prepared in the state $\ket{+_0}$ then for the register qubits in the states $\ket{q_1}$ and $\ket{q_2}$ respectively, the action on the ancilla from the first two gates is
\[
X^{q_2}Z^{q_1} \ket{+_0} = (-1)^{q_1q_2} \ket{+_{q_1}},
\]
which may be confirmed from Eq.~(\ref{Eq-conjP}). As we have already seen, the remaining two gates in Fig.~\ref{Fig_qubit-phase-gate} may be used to disentangle the ancilla from the register. However, instead it may be disentangled by a measurement that is specifically chosen to \emph{reveal no information} about the state of the register qubits.  A suitable choice is a measurement which projects onto the computational basis states $\ket{0}$ and $\ket{1}$ with measurement outcomes $0$ and $1$ respectively, termed a \emph{computational basis measurement}. This reveal no information about the register because the outcomes $0$ and $1$ each have a probability of a half regardless of the values of $q_1$ and $q_2$. For measurement outcome $m$, the action on the register qubits is
\[
 (-1)^{q_1q_2} \ket{+_{q_1}} \to (-1)^{q_1m} (-1)^{q_1q_2},
 \]
found by applying $\bra{m}$, using the relation $\expect{m|q} = (-1)^{qm}/\sqrt{2}$ and renormalising. The $(-1)^{q_1q_2}$ phase creates a $CZ$ exactly as with the unitary geometric phase method, however there is now an additional phase as though a $Z(m)$ gate has also been applied to the first qubit ($Z(m)\ket{q}=(-1)^{mq}\ket{q}$ for qubits). Because the value of $m$ is known, local controls may then be used to correct this error by applying $Z(-m)$ to the first qubit.
\newline
\indent
This technique may be used to replace the `disentangling' stage of all the gate methods considered so far with a measurement, and is valid with all the combinations of variable types for the register and ancilla we have considered. For any variable, a computational basis measurement is simply one which projects the ancilla onto $\ket{m}$ with measurement outcome $m$. The measurement always creates an additional error gate (or gates) and when the register and ancilla variable types match this is $Z(m)$, but in other cases it is not a Pauli gate and this is important later.\footnote{In general the error gate is $R(2m\pi /d )$ where $d$ is the dimension of the ancilla.} 
In all cases, local controls may be used to remove this error. There are many variations on this basic measurement-induced gate method, including using different preparation states, interactions, and measurements, e.g., see \cite{spiller2006quantum,nemoto2005universal}, although all gates work on the principles outlined above. A further closely related technique is indirect measurements of photons (quantum non-demolition detectors)~\cite{spiller2006quantum}. 

\subsection{Implementing gates using minimal control \label{sec:minimal}}

\begin{figure}[h!]
\begin{center}
\includegraphics[width=4.7cm]{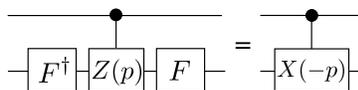}
\caption{A controlled $X$ gate can be generated using Fourier transforms and a controlled $Z$ gate.} \label{Fig_simulating-controlled-X}
\end{center}
\end{figure}

\begin{figure}[h!]
\center
\includegraphics[width=7cm]{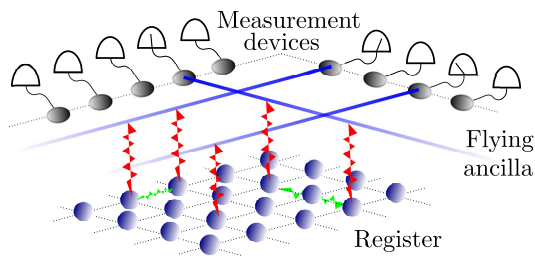}
\caption{Ancilla-based gate schemes may be used to implement universal quantum computation on a register using \emph{only} a single fixed interaction gate (red interactions) in conjunction with ancilla-preparation \cite{proctor2014quantum,Proctor2015ancilla} and in some cases aided by measurments of the ancillas  \cite{anders2010ancilla,Proctor2015ancilla,Korolkova2015cont}.} \label{Fig_minimal-idea}
\end{figure}

\begin{figure}[h!]
\begin{center}
\subfigure[]{
\resizebox*{2.3cm}{!}{\includegraphics{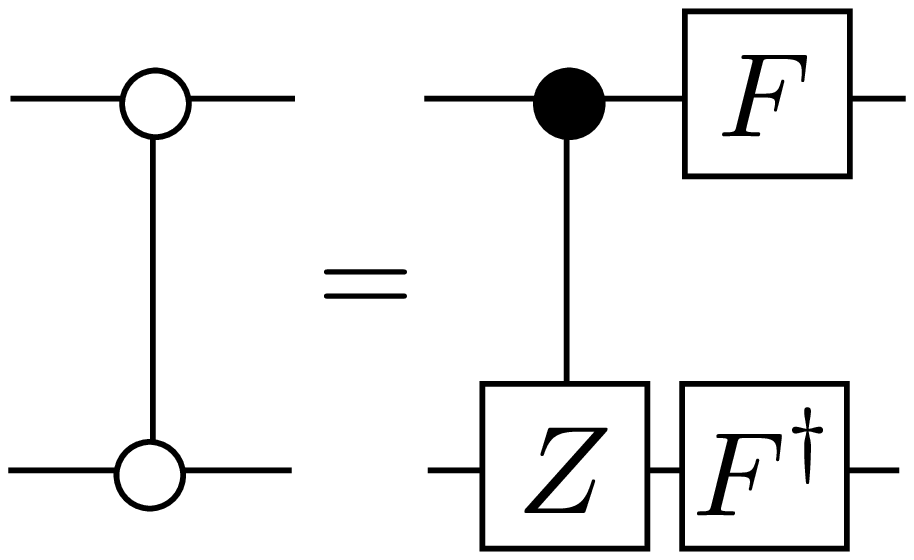}}}\hspace{20pt}
\subfigure[]{
\resizebox*{3.9cm}{!}{\includegraphics{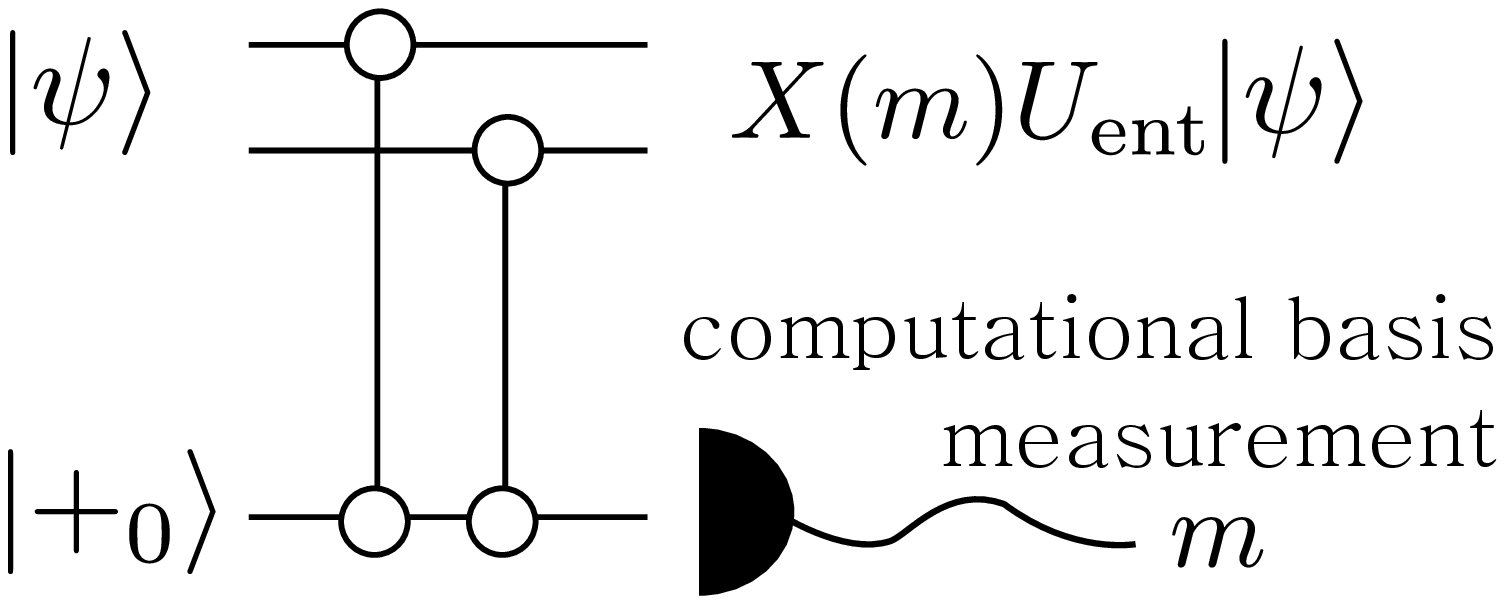}}}\hspace{20pt}
\subfigure[]{
\resizebox*{3.6cm}{!}{\includegraphics{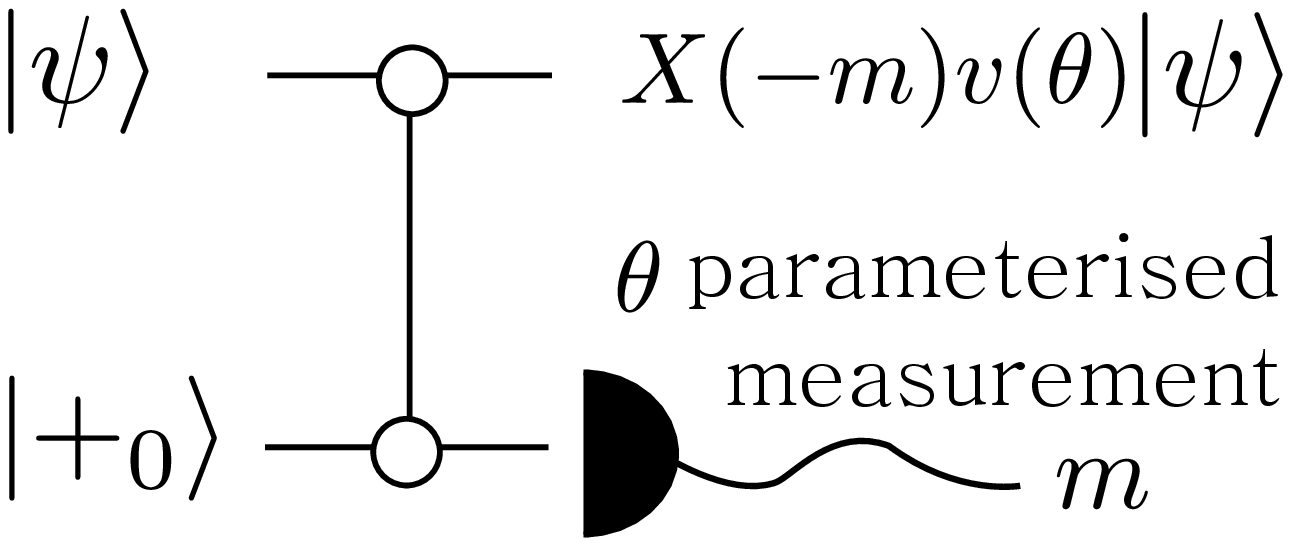}}}
\caption{Universal quantum computation is achieved in the ancilla-driven model using only a fixed ancilla-register interaction gate and measurements of the ancilla using two gate methods. (a) The form of the fixed interaction, decomposed into local and entangling gates. (b) An entangling gate is implemented up to a Pauli error via two interactions and a measurement (specifically, $U_{\text{ent}}=(F \otimes F)\cdot CZ$). (c) A universal set of local gates may be implemented up to a Pauli error via an interaction and measurement which depends on the desired gate. The form of the measurment and the implemented gate $v(\theta)$ are specified for qubits in the main text.} \label{Fig_ADQC-gates}
\end{center}
\end{figure}

\begin{figure}[h!]
\begin{center}
\includegraphics[width=7cm]{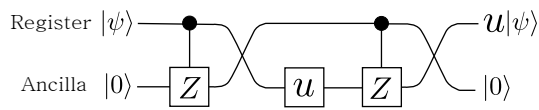}
\caption{An ancilla-register $\textsc{swap} \cdot CZ$ interaction gate (or the more general $\textsc{swap} \cdot CR(\theta)$) may be used to implement universal quantum computation in conjunction with local gates on the ancilla. Here the circuit shows how a local gate may be applied to a register subsystem by swapping the register subsystem into the ancilla, applying the gate, and swapping it back out. This is because $CZ$ acts as the identity when either system is in the $\ket{0}$ state. The crossing wires represent $\textsc{swap}$.}
\label{Fig_swap-gates}
\end{center}
\end{figure}

From a physical perspective, the implementation of a gate between an ancilla and register subsystem pair is achieved by evolving them via some interaction Hamiltonian $\hat{H}_{\text{int}}$. In some cases, applying this interaction for a varying length of time may well be possible, allowing a continuously parameterised family of interaction gates $U(t) = e^{-i \hat{H}_{\text{int}} t/\hbar}$. However, to implement additional interactions which are not of this form would require the engineering of more than one basic interaction (e.g, via varying a parameter in the Hamiltonian) and in general this may not be possible or may substantially complicate an experiment. For this reason, gate methods which require only a single fixed interaction Hamiltonian are preferable. Notably, this is the case with the geometric phase gate, and the measurement-based adaption discussed above. This is because $CZ(p)$ gates with different $p$ values can be achieved with one physical interaction via different interaction times.  Furthermore, $CX(p)$ gates can be generated via $CZ(p)$ and local Fourier transforms as shown in Fig.~\ref{Fig_simulating-controlled-X}. This may be easily confirmed from the relation $X(-p)=FZ(p)F^{\dagger}$ and the definition of a controlled gate.
%
%
\newline
\indent
It is interesting to consider whether the required physical controls can be limited still further. In general, access to individual ancilla and register subsystems to apply local gates may not be straightforward or even possible, e.g., in scatting-based interactions \cite{ciccarello2008extraction}, and limiting access to the register may be useful for further reducing sources of decoherence. Motivated by these ideas, methods to implement universal quantum computation on a register using \emph{only} a single register-ancilla interaction gate along with ancilla-preparation  \cite{proctor2014quantum,Proctor2015ancilla} and in some cases ancilla measurement \cite{anders2010ancilla,Proctor2015ancilla,Korolkova2015cont} have been developed. The idea behind these models is summarised in Fig.~\ref{Fig_minimal-idea}, and these gate methods are now explained. 
%
\newline
\indent
We first consider the \emph{ancilla-driven quantum computation} model which was introduced by Anders \emph{et al.} \cite{anders2010ancilla} for qubit computation. Recently, this has been extended to qudits \cite{Proctor2015ancilla} and QCVs \cite{Korolkova2015cont}. Interestingly, this model requires the ancilla and register systems to consist of the same variable type and size (same $d$ for qudits). As the idea is to implement universal quantum computation using only one fixed unitary gate, it is clear that this must be carefully chosen. The simplest suitable choice is a controlled-$Z$ gate followed by an $F$ and $F^{\dagger}$ gate on the register and ancilla subsystems respectively, as shown in Fig.~\ref{Fig_ADQC-gates} (a). (The three elementary gates are combined into a single two-system gate.)  Interacting two register systems in turn with the ancilla using this gate, followed by a computational basis measurement of the ancilla, implements an entangling gate on the two register systems as shown in Fig.~\ref{Fig_ADQC-gates} (b). By considering the effect of this sequence on the ancilla, it can be confirmed that this is essentially identical to the measured adaption of the geometric phase gate considered in Section~\ref{Sec_measured-gates}. As before, the measurement creates an error gate on the register, which is a Pauli gate as the ancilla and register are of the same variable type.
%
%
\newline
\indent
The addition of a method for implementing a universal family of local gates (i.e., a set of local gates that can generate any local gate) on any register subsystems is sufficient to make this scheme a universal quantum computer. Applying a gate from such a set on a register subsystem is achieved by an interaction with an ancilla followed by measuring the ancilla, with the measurement fixed by which gate from the family is to be implemented. Again, the desired gate is implemented only up to a Pauli error. This gate method is shown schematically in Fig~. \ref{Fig_ADQC-gates} (c). To be more explicit, we briefly consider the details of how this is achieved in the qubit case. For the register qubit in an arbitrary state $\alpha\ket{0} +  \beta \ket{1} $ the fixed interaction gate maps
\[
(\alpha\ket{0} +  \beta \ket{1})\ket{+_0} \to \alpha \ket{+_0} \ket{0} +  \beta \ket{+_1} \ket{1}.
\]
By performing a measurement which projects the ancilla onto $\ket{\theta_m} = R(-\theta)\ket{+_m}$ with outcomes $m=0$ or $m=1$ the gate $v(\theta)=HR(\theta)$ is implemented on the register qubit (followed by a Pauli error), which may easily be confirmed by applying $\bra{\theta_m}$ to the above expression. By varying $\theta$, the $v(\theta)$ gates are well-known to be sufficient to generate any single-qubit gate \cite{nielsen2010quantum}. The generalisation to other variable types is straightforward and simply requires a natural generalisation of $R(\theta)$, see \cite{Proctor2015ancilla,Korolkova2015cont}, but for brevity we do not review it here.
\newline
\indent
The measurement-induced errors might initially appear to be problematic as local controls have been assumed to be unavailable. However, these error gates can be commuted through to the end of the computation by changing some of the measurement basis parameters depending on preceding measurement outcomes along with some basic classical computations. The errors may then be absorbed into the final measurements of the register. This technique is often called \emph{classical feedforward} and is used in cluster-state quantum computing \cite{browne2006one}. It works only because the errors are Pauli gates and is most easily understood in terms of the Clifford group (the entangling gate here is a Clifford gate) which is beyond the scope of this review. The classical computation required to keep track of the corrections is simple but non-trivial in how it contributes to the overall computation~\cite{stepney2012framework}. 
\newline
\indent
The ancilla-driven model has been intensely studied recently, including: an analysis of gate errors obtained from inaccurate measurements \cite{morimae2010entanglement}; a investigation of the suitability of this model for secure quantum computation \cite{sueki2013ancilla}; and adaptions to allow for both a greater range of interactions \cite{shah2013ancilla}, and fixed measurements \cite{Proctor2015ancilla,halil2014minimum} which produce schemes which are universal only using probabilistic so-called \emph{repeat-until-success} gate implementations. An interesting question is whether an interaction locally equivalent to $CZ$ is the only gate that may be used to implement ancilla-driven quantum computation. It turns out that it is not: the alternative is based on a $\textsc{swap}$ gate, as introduced in Eq.~(\ref{eq:swap}). Specifically, it utilises the interaction $\textsc{swap} \cdot CZ$, and requires only minor changes to the model \cite{kashefi2009twisted,anders2010ancilla,Proctor2015ancilla}. 
\newline
\indent
However, there is a perhaps more interesting way to implement quantum computation using ancillas and this swap-type interaction, by returning to globally unitary dynamics as were used in the geometric phase gates. The action of $\textsc{swap}\cdot CZ$ on the computational basis states is
\[
  \ket{q_1}\ket{q_2} \xrightarrow{\textsc{swap} \cdot CZ} \omega^{q_1q_2} \ket{q_2} \ket{q_1},
\]
and hence if the ancilla (or the register) subsystem is in the $\ket{0}$ state it simply acts as a $\textsc{swap}$ gate. By interacting with a second register subsystem, this then implements an entangling gate between the logical register subsystem residing in the ancilla and this second register subsystem (the gate is $\textsc{swap}\cdot CZ$). A second interaction of the ancilla with the first register subsystem simply swaps the logical information back into the register and the overall effect is the application of $\textsc{swap}\cdot CZ$ to the two register systems \cite{proctor2013universal,proctor2014quantum}. Local gates may easily be applied to a register subsystem by swapping it into the ancilla, applying the gate, and swapping it back out \cite{proctor2013universal} as shown in Fig.~\ref{Fig_swap-gates}. Conveniently, the same methods hold with a less restricted $\textsc{swap} \cdot CR(\theta)$ interaction.
 %
%
\newline
\indent
The gate method for local gate on the register with the swap-type interaction described above has reverted to using unitary local controls (here of the ancilla). If we do not wish to rely on such controls, then with a minor adaption to the interaction it is possible to replace them with ancilla preparation in the different computational basis states, which then determines which local gates are applied to the register subsystem \cite{proctor2014minimal,Proctor2015ancilla}. For qubits, this is particularly simple: by appending an additional (fixed) local $u$ gate on the ancilla to the interaction (after $\textsc{swap}\cdot CR(\theta)$), two interactions of a register qubit with an ancilla in the state $\ket{0}$ or $\ket{1}$ respectively implement $u$ and $R(\theta)uR(\theta)$ on the register qubit respectively \cite{proctor2014minimal}. For qubits there are many choices such that this is a universal set for single-qubit gates (e.g., $u=H$, $\theta=\pi/4$). This therefore provides a scheme for doing universal quantum computing, where the program is simply defined by the ordering of the fixed-type interactions between the ancillas and different register subsystems, and by the initial states of the ancillas.
 \newline
 \indent
Finally, we note that the ancilla-based gate schemes outlined throughout this section have wider applications than just universal quantum computation, where they are very useful as building blocks for practical quantum computer architectures \cite{Nickerson2014nqit}. They can also be adapted to prepare exotic quantum states, which are used as resources for many purposes in quantum information (e.g., quantum cryptography \cite{hughes1995quantum} or quantum sensing networks \cite{komar2014quantum}), and are interesting from a fundamental perspective. For example, a qudit ancilla may be used to create highly entangled `$W$' and GHZ states using a similar construction \cite{ionicioiu2008generalized} to the efficient Toffoli gate in Fig.~\ref{Fig_ancilla-Toffoli-gate}, and so-called `cat states' of a QCV may be created using a qubit ancilla \cite{van2011optical}.

\section{Proposals and Implementations \label{sec:expt}}

\begin{figure}[h!]
\begin{center}
\includegraphics[width=8cm]{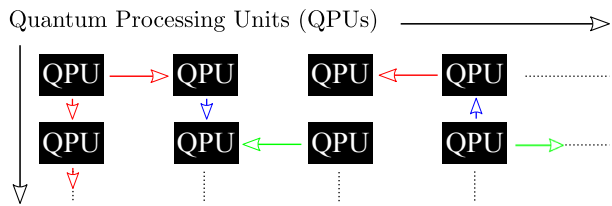}
\caption{Many proposals for universal, scalable and fault-tolerant quantum computer utilise fixed-sized registers, or quantum processing units (QPUs), entangled via ancilla or quantum communication buses \cite{Nickerson2014nqit,monroe2014large}. Here ancilla sent at different times are denoted by different coloured arrows.} \label{Fig_bus_architecture}
\end{center}
\end{figure}

Future designs for a universal, scalable and fault-tolerant quantum computer will likely be based around modular processing units entangled via ancillary systems, as illustrated in Fig.~\ref{Fig_bus_architecture}. A recent proposal of this type \cite{Nickerson2014nqit} forms the core objective of the Networked Quantum Information Technologies (NQIT) Quantum Technology Hub lead by Oxford University. NQIT aims to construct a scalable network of high fidelity quantum registers linked via more lossy optical ancillas. There are a variety of qubit register implementations that are suitable for this architecture, with one of the most advanced being ion trap technology \cite{monroe2014large}. Promising recent experimental progress has been made in this direction \cite{hucul2014modular}.  
%
%
\newline
\indent
There are a wide range of basic gate implementations that employ ancillary systems. Quantum gates have been implemented by using photonic ancillas to entangle spin \cite{carter2013quantum,luxmoore2013interfacing} or atomic \cite{reiserer2014quantum,tiecke2014nanophotonic} qubits. Superconducting flux qubits are usually entangled via a transmission line resonator bus \cite{stern2014flux}. They have also been coupled to an ensemble of nitrogen-vacancy (NV) centres in diamond encoding a qubit \cite{zhu2011coherent}. This may be suitable for linking a register of such NV ensemble-based qubits using ancillary flux qubits \cite{qiu2014coupling}, with a related proposal in \cite{lu2013quantum}. This illustrates an obvious but important point, that any quantum system engineered for useful information processing properties can be considered for both register and ancilla roles, giving a very wide range of options for designing such devices.  Spin-ensembles could serve as a long-coherence time register (often called a quantum memory) as such ensemble are naturally resilient to various forms of loss and decoherence~\cite{rabl2006hybrid}.
\newline
\indent
These examples are all implemented or proposed as qubit-based architectures with qubit or QCV ancillas, but many have the potential to support qudit encodings as well, e.g., atoms have many energy levels and harmonic oscillators may encode qudits. The theoretical advantages of adopting such qudit-based platforms were discussed in Section~\ref{sec:hybrid}, and further incentive to explore higher-dimensional protocols is provided by a range of impressive experiments demonstrating quantum control of qudits, including superconducting \cite{neeley2009emulation}, atomic \cite{smith2013quantum} and photonic \cite{bent2015experimental,lima2011experimental} systems.
 \newline
 \indent
Finally, to relate this discussion to some concrete physics, we consider one particular Hamiltonian: the Jaynes-Cummings model \cite{jaynes1963comparison}, which is relevant for a large range of architectures for quantum computing. This is a canonical scheme which describes the coupling of a qubit to a harmonic oscillator. Mathematically, the Jaynes-Cummings Hamiltonian is given by
  \[
  \hat{H}_{\textsc{jc}}=\underbrace{ \omega\hat{a}^\dagger \hat{a}}_{\text{oscillator}}+ \underbrace{ \omega_0 \frac{Z}{2}}_{\text{qubit}}+\underbrace{g(\hat{a}^\dagger \sigma_- +\hat{a}\sigma_+)}_{\text{interaction}},
  \]
where $\sigma_{+} =  \ket{0}\bra{1}$ and  $\sigma_{-} =  \ket{1}\bra{0}$; $\hat{a}^{\dagger}$ and $\hat{a}$ are the creation and annihilation operators for a harmonic oscillator (related to $\hat{x}$ and $\hat{p}$ via $ \hat{a}^{\dagger}=(\hat{x}-i\hat{p})/\sqrt{2}$ and $ \hat{a}=(\hat{x}+i\hat{p})/\sqrt{2}$); $\omega $ is the frequency of the harmonic oscillator; $ \omega_0 $ is the frequency of the qubit and $ g $ represents the qubit-oscillator coupling strength. The Jaynes-Cummings model finds application in describing a wide range of experimental systems including atom-cavity coupling \cite{shore1993jaynes}, circuit QED \cite{blais2004cavity} and for a variety of qubit types coupling to mechanical oscillators \cite{wallquist2009hybrid,groblacher2009observation}. A harmonic oscillator naturally lends itself to a QCV encoding, but by encoding a qudit as the first $d$ energy eigenstates of a harmonic oscillator, the Jaynes-Cummings model can also be used to describe a qudit-qubit coupling \cite{mischuck2013qudit}. In the \emph{dispersive limit} ($g/(\omega_0 - \omega) \ll 1$) this interaction is approximately of the form $\hat{H}_{\text{eff}} = \sigma_z  \hat{a}^{\dagger}\hat{a}$ \cite{blais2004cavity} (up to local terms) which with the QCV encoding may generate qubit-controlled \emph{phase-space rotations}. These are an alternative type of ancilla-register gate to the controlled Pauli operators considered throughout Section~\ref{sec:ancilla}, for which a variety of gate methods closely related to those discussed here have been developed, see e.g., \cite{spiller2006quantum,louis2007efficiencies}. However, this interaction may be easily transformed into a controlled Pauli gate \cite{van2008hybrid,wang2002simulation}, and hence is also suitable for the gate methods reviewed in Section~\ref{sec:ancilla}. A qudit-qudit or qudit-QCV hybrid device would require an alternative to the Jaynes-Cummings model, for example the \emph{generalised} Jaynes-Cummings model, which describes the coupling of a many-level particle coupled to a harmonic oscillator \cite{ionicioiu2009generalized}, or coupled harmonic oscillators.

\section{Summary \label{sec:conclude}}
Binary encoding has become the standard for classical digital computers, but there is no \emph{a priori} reason why binary should also be the best choice for quantum computers. Computation can equally well be carried out in any convenient basis, and with many quantum systems being higher dimensional, it is well worth asking which is the natural basis to use for each possible type of quantum hardware. There are some constraints: Blume-Kohout \emph{et al.} \cite{blume02a} show that the dimension of the subsystems must not grow faster than logarithmically with the size of the problem in order for the quantum computer to be efficient and scalable, but for practical hybrid systems, subsystem scaling will not be a significant issue. 
 \newline
 \indent
One good reason for the ubiquity of binary encoding in classical computation is undoubtedly simplicity. However, as we have seen in Section~\ref{sec:hybrid}, for quantum systems the same formalisms carries over to higher dimensions with few changes: qudits and QCVs are not significantly harder to handle theoretically.  
There are advantages and trade-offs to using qudits instead of qubits: better performance of error correction \cite{watson2015qudit,campbell2014enhanced,anwar2014fast,campbell2012magic,andrist2015error,duclos2013kitaev}, fewer gates  \cite{Muthukrishnan2000Multivalued,stroud2002quantum}, and more robust algorithms \cite{parasa2011quantum,zilic2007scaling,parasa2012quantum} are set against more complex gate operations and measurements.  The problem itself may be more naturally encoded in qudits, or QCVs if the problems has continuous parameters, and this may lead to a simpler algorithm \cite{travaglione2001generation}.  Furthermore, using a heterogeneous architecture in which both qudits and QCVs are available for the same computation allows different components of the problem to be optimally encoded \cite{lloyd2003hybrid}.
 \newline
 \indent
Using ancillas to enact qubit quantum gates was established in some of the earliest proposed implementations of quantum computation \cite{cirac1995quantum}, and from the start the ancillas were of very different types to the register qubits. The formalism described in Section~\ref{sec:hybrid} allows ancilla-based quantum gates to be described in a similar framework for all dimensions of register and ancillas, as was presented in Section~\ref{sec:ancilla}. To simplify the experimental requirements, schemes with a single, fixed interaction between register and ancillas are the most useful. This minimises the disturbance to the register qubits, preserving quantum coherence for longer. It is possible to minimise the interactions to the point where universal quantum computation can be performed by a sequence of ancilla-register interactions plus preparation of the ancillas in different computational basis states~\cite{proctor2014minimal,Proctor2015ancilla}.  
 \newline
 \indent
The tools presented here for describing quantum computation using ancillas to enact the basic quantum gates allow quantum engineers to take a fresh look at the possibilities for their preferred quantum systems, to choose combinations that optimise their best properties and hence deliver a scalable universal quantum computer with fewer quantum resources.


\section*{Acknowledgments}
This work was partially supported by the EPSRC under Grant EP/L022303/1.  TJP is funded by a University of Leeds Scholarship. Discussions with many colleagues helped to develop our ideas in this review, but special thanks go to Erika Andersson for introducing us to the topic of ancilla-mediated quantum gates, and to Joschka Roffe for assistance with Section~\ref{sec:expt}, and thoroughly checking the manuscript and references.

\bibliographystyle{tCPH}
\bibliography{Bib_Library}

\begin{thebibliography}{100}
\newcommand{\noopsort}[1]{}
\newcommand{\printfirst}[2]{#1}
\newcommand{\singleletter}[1]{#1}
\newcommand{\switchargs}[2]{#2#1}
\providecommand{\url}[1]{\normalfont{#1}}
\providecommand{\urlprefix}{Available at }

\bibitem{shor1994algorithms}
P.W. Shor, \emph{Algorithms for quantum computation: Discrete logarithms and
  factoring}, in \emph{Foundations of Computer Science, 1994 Proceedings., 35th
  Annual Symposium on}, 1994, pp. 124--134.

\bibitem{shor1997polynomial}
P.W. Shor, \emph{Polynomial-time algorithms for prime factorization and
  discrete logarithms on a quantum computer}, SIAM journal on computing 26
  (1997), pp. 1484--1509.

\bibitem{rivest1978method}
R.L. Rivest, A. Shamir, and L. Adleman, \emph{A method for obtaining digital
  signatures and public-key cryptosystems}, Communications of the ACM 21
  (1978), pp. 120--126.

\bibitem{grover1996fast}
L.K. Grover, \emph{A fast quantum mechanical algorithm for database search}, in
  \emph{Proceedings of the twenty-eighth annual ACM symposium on Theory of
  computing}, 1996, pp. 212--219.

\bibitem{schuld2015introduction}
M. Schuld, I. Sinayskiy, and F. Petruccione, \emph{An introduction to quantum
  machine learning}, Contemporary Physics 56 (2015), pp. 172--185.

\bibitem{brown2010using}
K.L. Brown, W.J. Munro, and V.M. Kendon, \emph{Using quantum computers for
  quantum simulation}, Entropy 12 (2010), pp. 2268--2307.

\bibitem{feynman1982simulating}
R.P. Feynman, \emph{Simulating physics with computers}, Int. J. Th. Phys. 21
  (1982), pp. 467--488.

\bibitem{aaronson2015read}
S. Aaronson, \emph{Read the fine print}, Nature Physics 11 (2015), pp.
  291--293.

\bibitem{aaronson2005guest}
S. Aaronson, \emph{{NP}-complete problems and physical reality}, ACM Sigact
  News 36 (2005), pp. 30--52.

\bibitem{bennett1997strengths}
C.H. Bennett, E. Bernstein, G. Brassard, and U. Vazirani, \emph{Strengths and
  weaknesses of quantum computing}, SIAM journal on Computing 26 (1997), pp.
  1510--1523.

\bibitem{bacon2010recent}
D. Bacon and W. Van~Dam, \emph{Recent progress in quantum algorithms},
  Communications of the ACM 53 (2010), pp. 84--93.

\bibitem{gottesman2010introduction}
D. Gottesman, \emph{An introduction to quantum error correction and
  fault-tolerant quantum computation}, ArXiv:0904.2557 68 (2010), pp. 13--60.

\bibitem{terhal2015quantum}
B.M. Terhal, \emph{Quantum error correction for quantum memories}, Rev. Mod.
  Phys. 87 (2015), p. 307.

\bibitem{zhong2015optically}
M. Zhong, M.P. Hedges, R.L. Ahlefeldt, J.G. Bartholomew, S.E. Beavan, S.M.
  Wittig, J.J. Longdell, and M.J. Sellars, \emph{Optically addressable nuclear
  spins in a solid with a six-hour coherence time}, Nature 517 (2015), pp.
  177--180.

\bibitem{wang2009coupling}
Y.D. Wang, A. Kemp, and K. Semba, \emph{Coupling superconducting flux qubits at
  optimal point via dynamic decoupling with the quantum bus}, Phys. Rev. B 79
  (2009), p. 024502.

\bibitem{xue2012fast}
Z.Y. Xue, \emph{Fast geometric gate operation of superconducting charge qubits
  in circuit qed}, Quantum Inf. Process. 11 (2012), pp. 1381--1388.

\bibitem{spiller2006quantum}
T.P. Spiller, K. Nemoto, S.L. Braunstein, W.J. Munro, P. van  Loock, and G.J.
  Milburn, \emph{Quantum computation by communication}, New J. Phys. 8 (2006),
  p.~30.

\bibitem{cirac1995quantum}
J.I. Cirac and P. Zoller, \emph{Quantum computations with cold trapped ions},
  Phys. Rev. Lett. 74 (1995), p. 4091.

\bibitem{epstein2012adiabatic}
C. Epstein, \emph{Adiabatic quantum computing: An overview}, Quantum Complexity
  Theory 6 (2012), p. 845.

\bibitem{das2008colloquium}
A. Das and B.K. Chakrabarti, \emph{Colloquium: Quantum annealing and analog
  quantum computation}, Rev. Mod. Phys. 80 (2008), p. 1061.

\bibitem{trummer2015multiple}
I. Trummer and C. Koch, \emph{Multiple query optimization on the {D-Wave 2X}
  adiabatic quantum computer}, arXiv preprint arXiv:1510.06437  (2015).

\bibitem{knuth1968art}
D.E. Knuth, \emph{The Art of Computer Programming}, Vol. 2: Seminumerical
  Algorithms (see page 190), 2nd ed., MA: Addison-Wesley, 1980.

\bibitem{weyl1950theory}
H. Weyl, \emph{The theory of groups and quantum mechanics}, Courier Dover
  Publications, 1950.

\bibitem{vourdas2004quantum}
A. Vourdas, \emph{Quantum systems with finite {H}ilbert space}, Rep. Prog.
  Phys. 67 (2004), p. 267.

\bibitem{wootters1987wigner}
W.K. Wootters, \emph{A wigner-function formulation of finite-state quantum
  mechanics}, Ann. Phys. 176 (1987), pp. 1--21.

\bibitem{gibbons2004discrete}
K.S. Gibbons, M.J. Hoffman, and W.K. Wootters, \emph{Discrete phase space based
  on finite fields}, Phys. Rev. A 70 (2004), p. 062101.

\bibitem{sylvester2012collected}
J.J. Sylvester and H.F. Baker, \emph{The collected mathematical papers of James
  Joseph Sylvester}, Vol.~3, Cambridge University Press, 2012.

\bibitem{shannon1941mathematical}
C.E. Shannon, \emph{Mathematical theory of the differential analyzer}, J. Math.
  Phys. MIT 20 (1941), pp. 337--354.

\bibitem{thomson1875mechanical}
W. Thomson, \emph{Mechanical integration of the general linear differential
  equation of any order with variable coefficients}, Proc. Roy. Soc. 24 (1875),
  pp. 271--275.

\bibitem{kendon2010quantum}
V.M. Kendon, K. Nemoto, and W.J. Munro, \emph{Quantum analogue computing},
  Phil. Trans. R. Soc. A 368 (2010), pp. 3609--3620.

\bibitem{braunstein2005quantum}
S.L. Braunstein and P. van  Loock, \emph{Quantum information with continuous
  variables}, Rev. Mod. Phys. 77 (2005), pp. 513--577.

\bibitem{raussendorf2012key}
R. Raussendorf, \emph{Key ideas in quantum error correction}, Phil. Trans. R.
  Soc. A 370 (2012), pp. 4541--4565.

\bibitem{gottesman1999heisenberg}
D. Gottesman, \emph{The Heisenberg representation of quantum computers}, in
  \emph{Proceedings of the XXII International Colloquium on Group Theoretical
  Methods in Physics}, International Press arXiv preprint quant-ph/9807006,
  1999, p.~32.

\bibitem{bartlett2002efficient}
S.D. Bartlett, B.C. Sanders, S.L. Braunstein, and K. Nemoto, \emph{Efficient
  classical simulation of continuous variable quantum information processes.},
  Phys. Rev. Lett. 88 (2002), p. 097904.

\bibitem{gottesman1999fault}
D. Gottesman, \emph{Fault-tolerant quantum computation with higher-dimensional
  systems}, in \emph{Quantum Computing and Quantum Communications}, Springer,
  1999, pp. 302--313.

\bibitem{browne2006one}
D.E. Browne and H.J. Briegel, \emph{One-way Quantum Computation}, in
  \emph{Lectures on Quantum Information}, chap. 5.3, Wiley-VCH, 2006, chap.
  5.3.

\bibitem{zhou2003quantum}
D.L. Zhou, B. Zeng, Z. Xu, and C.P. Sun, \emph{Quantum computation based on
  d-level cluster state}, Phys. Rev. A 68 (2003), p. 062303.

\bibitem{menicucci2006universal}
N.C. Menicucci, P. van  Loock, M. Gu, C. Weedbrook, T.C. Ralph, and M.A.
  Nielsen, \emph{Universal quantum computation with continuous-variable cluster
  states}, Phys. Rev. Lett. 97 (2006), p. 110501.

\bibitem{van2013efficient}
M. Van~den  Nest, \emph{Efficient classical simulations of quantum {F}ourier
  transforms and normalizer circuits over abelian groups}, Quant. Info. Comput.
  13 (2013), pp. 1007--1037.

\bibitem{poot2012mechanical}
M. Poot and H.S.J. van~der  Zant, \emph{Mechanical systems in the quantum
  regime}, Physics Reports 511 (2012), pp. 273--335.

\bibitem{gerry2005introductory}
C. Gerry and P. Knight, \emph{Introductory quantum optics}, Cambridge
  university press, 2005.

\bibitem{radmore1997methods}
P.M. Radmore and S.M. Barnett, \emph{Methods in theoretical quantum optics},
  Oxford University Press Oxford,, UK, 1997.

\bibitem{bennett93teleporting}
C.H. Bennett, G. Brassard, C. Cr{\'e}peau, R. Jozsa, A. Peres, and W.K.
  Wootters, \emph{Teleporting an unknown quantum state via dual classical and
  einstein-podolsky-rosen channels}, Phys. Rev. Lett. 70 (1993), pp.
  1895--1899.

\bibitem{silberhorn2007detecting}
C. Silberhorn, \emph{Detecting quantum light}, Contemporary Physics 48 (2007),
  pp. 143--156.

\bibitem{leonhardt2005measuring}
U. Leonhardt, \emph{Measuring the quantum state of light}, Vol.~1, Cambridge
  University Press, 2005.

\bibitem{hughes1995quantum}
R.J. Hughes, D. Alde, P. Dyer, G. Luther, G. Morgan, and M. Schauer,
  \emph{Quantum cryptography}, Contemporary Physics 36 (1995), pp. 149--163.

\bibitem{einstein1935can}
A. Einstein, B. Podolsky, and N. Rosen, \emph{Can quantum-mechanical
  description of physical reality be considered complete?}, Phys. Rev. 47
  (1935), p. 777.

\bibitem{brylinski2002universal}
J.L. Brylinski and R. Brylinski, \emph{Universal quantum gates}, Chapman \&
  Hall / CRC Press, 2002.

\bibitem{Lloyd1999quantum}
S. Lloyd and S.L. Braunstein, \emph{Quantum computation over continuous
  variables}, Phys. Rev. Lett. 82 (1999), pp. 1784--1787.

\bibitem{hostens2005stabilizer}
E. Hostens, J. Dehaene, and B. De~Moor, \emph{Stabilizer states and clifford
  operations for systems of arbitrary dimensions and modular arithmetic}, Phys.
  Rev. A 71 (2005), p. 042315.

\bibitem{Proctor2015ancilla}
T.J. Proctor and V. Kendon, \emph{Higher-dimensional ancilla-driven quantum
  computation}, arXiv:1510.06462  (2015).

\bibitem{howard2012qudit}
M. Howard and J. Vala, \emph{Qudit versions of the qubit $\pi$/8 gate}, Phys.
  Rev. A 86 (2012), p. 022316.

\bibitem{Muthukrishnan2000Multivalued}
A. Muthukrishnan and C.R. Stroud~Jr, \emph{Multivalued logic gates for quantum
  computation}, Phys. Rev. A 62 (2000), p. 052309.

\bibitem{stroud2002quantum}
A. Stroud and C.R. Muthukrishnan, \emph{Quantum fast {F}ourier transform using
  multilevel atoms}, J. Mod. Opt. 49 (2002), pp. 2115--2127.

\bibitem{parasa2011quantum}
V. Parasa and M. Perkowski, \emph{Quantum phase estimation using multivalued
  logic}, in \emph{Multiple-Valued Logic (ISMVL), 2011 41st IEEE International
  Symposium on}, 2011, pp. 224--229.

\bibitem{zilic2007scaling}
Z. Zilic and K. Radecka, \emph{Scaling and better approximating quantum fourier
  transform by higher radices}, IEEE Transactions on computers 56 (2007), pp.
  202--207.

\bibitem{parasa2012quantum}
V. Parasa and M. Perkowski, \emph{Quantum Pseudo-Fractional {F}ourier Transform
  Using Multiple-Valued Logic}, in \emph{Multiple-Valued Logic (ISMVL), 2012
  42nd IEEE International Symposium on}, 2012, pp. 311--314.

\bibitem{watson2015qudit}
F.H.E. Watson, E.T. Campbell, H. Anwar, and D.E. Browne, \emph{Qudit color
  codes and gauge color codes in all spatial dimensions}, Phys. Rev. A 92
  (2015), p. 022312.

\bibitem{campbell2014enhanced}
E.T. Campbell, \emph{Enhanced fault-tolerant quantum computing in d-level
  systems}, Phys. Rev. Lett. 113 (2014), p. 230501.

\bibitem{anwar2014fast}
H. Anwar, B.J. Brown, E.T. Campbell, and D.E. Browne, \emph{Fast decoders for
  qudit topological codes}, New J. Phys. 16 (2014), p. 063038.

\bibitem{campbell2012magic}
E.T. Campbell, H. Anwar, and D.E. Browne, \emph{Magic-state distillation in all
  prime dimensions using quantum reed-muller codes}, Phys. Rev. X 2 (2012), p.
  041021.

\bibitem{andrist2015error}
R.S. Andrist, J.R. Wootton, and H.G. Katzgraber, \emph{Error thresholds for
  abelian quantum double models: Increasing the bit-flip stability of
  topological quantum memory}, Phys. Rev. A 91 (2015), p. 042331.

\bibitem{duclos2013kitaev}
G. Duclos-Cianci and D. Poulin, \emph{Kitaev's $\mathbb{Z}_d$-code threshold
  estimates}, Phys. Rev. A 87 (2013), p. 062338.

\bibitem{lloyd2003hybrid}
S. Lloyd, \emph{Hybrid quantum computing}, arXiv:quant-ph/0008057  (2003).

\bibitem{travaglione2001generation}
B.C. Travaglione and G.J. Milburn, \emph{Generation of eigenstates using the
  phase-estimation algorithm}, Phys. Rev. A 63 (2001), p. 032301.

\bibitem{munro2005weak}
W.J. Munro, K. Nemoto, and T.P. Spiller, \emph{Weak nonlinearities: a new route
  to optical quantum computation}, New J. Phys. 7 (2005), p. 137.

\bibitem{armour2002entanglement}
A.D. Armour, M.P. Blencowe, and K.C. Schwab, \emph{Entanglement and decoherence
  of a micromechanical resonator via coupling to a cooper-pair box}, Phys. Rev.
  Lett. 88 (2002), pp. 148301--148301.

\bibitem{milburn1999simulating}
G.J. Milburn, \emph{Simulating nonlinear spin models in an ion trap},
  arXiv:quant-ph/9908037  (1999).

\bibitem{van2008hybrid}
P. Van~Loock, W.J. Munro, K. Nemoto, T.P. Spiller, T.D. Ladd, S.L. Braunstein,
  and G.J. Milburn, \emph{Hybrid quantum computation in quantum optics}, Phys.
  Rev. A 78 (2008), p. 022303.

\bibitem{proctor2014quantum}
T.J. Proctor, S. Dooley, and V. Kendon, \emph{Quantum computation mediated by
  ancillary qudits and spin coherent states}, Phys. Rev. A 91 (2015), p.
  012308.

\bibitem{milburn2000ion}
G. Milburn, S. Schneider, and D. James, \emph{Ion trap quantum computing with
  warm ions}, Fortschritte der Physik 48 (2000), pp. 801--810.

\bibitem{louis2008loss}
S.G.R. Louis, W.J. Munro, T.P. Spiller, and K. Nemoto, \emph{Loss in hybrid
  qubit-bus couplings and gates}, Phys. Rev. A 78 (2008), p. 022326.

\bibitem{wang2002simulation}
X. Wang and P. Zanardi, \emph{Simulation of many-body interactions by
  conditional geometric phases}, Phys. Rev. A 65 (2002), p. 032327.

\bibitem{louis2007efficiencies}
S.G.R. Louis, K. Nemoto, W.J. Munro, and T.P. Spiller, \emph{The efficiencies
  of generating cluster states with weak nonlinearities}, New J. Phys. 9
  (2007), p. 193.

\bibitem{brown2011ancilla}
K.L. Brown, S. De, V.M. Kendon, and W.J. Munro, \emph{Ancilla-based quantum
  simulation}, New J. Phys. 13 (2011), p. 095007.

\bibitem{horsman2011reduce}
C. Horsman, K.L. Brown, W.J. Munro, and V.M. Kendon, \emph{Reduce, reuse,
  recycle for robust cluster-state generation}, Phys. Rev. A 83 (2011), p.
  042327.

\bibitem{khosla2013quantum}
K. Khosla, M. Vanner, W. Bowen, and G. Milburn, \emph{Quantum state preparation
  of a mechanical resonator using an optomechanical geometric phase}, New J.
  Phys. 15 (2013), p. 043025.

\bibitem{brown2012layer}
K.L. Brown, C. Horsman, and W.J. Kendon V.and~Munro, \emph{Layer-by-layer
  generation of cluster states}, Phys. Rev. A 85 (2012), p. 052305.

\bibitem{ionicioiu2009generalized}
R. Ionicioiu, T.P. Spiller, and W.J. Munro, \emph{Generalized toffoli gates
  using qudit catalysis}, Phys. Rev. A 80 (2009), p. 012312.

\bibitem{gottesman1997stabilizer}
D. Gottesman, \emph{Stabilizer codes and quantum error correction}, arXiv
  preprint quant-ph/9705052  (1997).

\bibitem{nielsen2010quantum}
M.A. Nielsen and I.L. Chuang, \emph{Quantum computation and quantum
  information}, Cambridge university press, 2010.

\bibitem{shi2002both}
Y. Shi, \emph{Both toffoli and controlled-not need little help to do universal
  quantum computation}, arXiv preprint quant-ph/0205115  (2002).

\bibitem{saeedi2013linear}
M. Saeedi and M. Pedram, \emph{Linear-depth quantum circuits for n-qubit
  toffoli gates with no ancilla}, Phys. Rev. A 87 (2013), p. 062318.

\bibitem{lanyon2009simplifying}
B.P. Lanyon, M. Barbieri, M.P. Almeida, T. Jennewein, T.C. Ralph, K.J. Resch,
  G.J. Pryde, J.L. O'Brien, A. Gilchrist, and A.G. White, \emph{Simplifying
  quantum logic using higher-dimensional {H}ilbert spaces}, Nature Phys. 5
  (2009), pp. 134--140.

\bibitem{nemoto2005universal}
K. Nemoto and W.J. Munro, \emph{Universal quantum computation on the power of
  quantum non-demolition measurements}, Phys. Lett. A 344 (2005), pp. 104--110.

\bibitem{ciccarello2008extraction}
F. Ciccarello, M. Paternostro, M.S. Kim, and G.M. Palma, \emph{Extraction of
  singlet states from noninteracting high-dimensional spins}, Phys. Rev. Lett.
  100 (2008), p. 150501.

\bibitem{anders2010ancilla}
J. Anders, D.K.L. Oi, E. Kashefi, D.E. Browne, and E. Andersson,
  \emph{Ancilla-driven universal quantum computation}, Phys. Rev. A 82 (2010),
  p. 020301(R).

\bibitem{Korolkova2015cont}
N. Korolkova, T. Nakano, and E. Andersson, \emph{Continuous-variable
  ancilla-driven quantum computation (in preparation)}, (In Preparation)
  (2015).

\bibitem{stepney2012framework}
S. Stepney, V. Kendon, P. Hines, and A. Sebald, \emph{A framework for heterotic
  computing}, EPTCS 95  (2012), pp. 263--273.

\bibitem{morimae2010entanglement}
T. Morimae and J. Kahn, \emph{Entanglement-fidelity relations for inaccurate
  ancilla-driven quantum computation}, Phys. Rev. A 82 (2010), p. 052314.

\bibitem{sueki2013ancilla}
T. Sueki, T. Koshiba, and T. Morimae, \emph{Ancilla-driven universal blind
  quantum computation}, Phys. Rev. A 87 (2013), p. 060301.

\bibitem{shah2013ancilla}
K. Halil-Shah and D.K.L. Oi, \emph{Ancilla Driven Quantum Computation with
  arbitrary entangling strength}, in \emph{Theory of Quantum Computation,
  Communication, and Cryptography, 8th Conference, TQC 2013, LIPIcs-Leibniz
  International Proceedings in Informatics, Vol. 23.}, 2013.

\bibitem{halil2014minimum}
K. Halil-Shah and D.K.L. Oi, \emph{A minimum control ancilla driven quantum
  computation scheme with repeat-until-success style gate generation}, arXiv
  preprint arXiv:1401.8004  (2014).

\bibitem{kashefi2009twisted}
E. Kashefi, D.K.L. Oi, D. Browne, J. Anders, and E. Andersson, \emph{Twisted
  graph states for ancilla-driven universal quantum computation}, Electronic
  Notes in Theoretical Computer Science 249 (2009), pp. 307--331.

\bibitem{proctor2013universal}
T.J. Proctor, E. Andersson, and V. Kendon, \emph{Universal quantum computation
  by the unitary control of ancilla qubits and using a fixed ancilla-register
  interaction}, Phys. Rev. A 88 (2013), p. 042330.

\bibitem{proctor2014minimal}
T.J. Proctor and V. Kendon, \emph{Minimal ancilla mediated quantum
  computation}, EPJ Quantum Technology, 1:13  (2014).

\bibitem{Nickerson2014nqit}
N.H. Nickerson, J.F. Fitzsimons, and S.C. Benjamin, \emph{Freely scalable
  quantum technologies using cells of 5-to-50 qubits with very lossy and noisy
  photonic links}, Phys. Rev. X 4 (2014), p. 041041.

\bibitem{komar2014quantum}
P. Komar, E.M. Kessler, M. Bishof, L. Jiang, A.S. S{\o}rensen, J. Ye, and M.D.
  Lukin, \emph{A quantum network of clocks}, Nature Physics 10 (2014), pp.
  582--587.

\bibitem{ionicioiu2008generalized}
R. Ionicioiu, A.E. Popescu, W.J. Munro, and T.P. Spiller, \emph{Generalized
  parity measurements}, Phys. Rev. A 78 (2008), p. 052326.

\bibitem{van2011optical}
P. van  Loock, \emph{Optical hybrid approaches to quantum information}, Laser
  Photon. Rev. 5 (2011), pp. 167--200.

\bibitem{monroe2014large}
C. Monroe, R. Raussendorf, A. Ruthven, K.R. Brown, P. Maunz, L.M. Duan, and J.
  Kim, \emph{Large-scale modular quantum-computer architecture with atomic
  memory and photonic interconnects}, Phys. Rev. A 89 (2014), p. 022317.

\bibitem{hucul2014modular}
D. Hucul, I.V. Inlek, G. Vittorini, C. Crocker, S. Debnath, S.M. Clark, and C.
  Monroe, \emph{Modular entanglement of atomic qubits using photons and
  phonons}, Nature Phys. 11 (2015), p.~37.

\bibitem{carter2013quantum}
S.G. Carter, T.M. Sweeney, M. Kim, C.S. Kim, D. Solenov, S.E. Economou, T.L.
  Reinecke, L. Yang, A.S. Bracker, and D. Gammon, \emph{Quantum control of a
  spin qubit coupled to a photonic crystal cavity}, Nature Photonics 7 (2013),
  pp. 329--334.

\bibitem{luxmoore2013interfacing}
I. Luxmoore, N. Wasley, A. Ramsay, A. Thijssen, R. Oulton, M. Hugues, S.
  Kasture, V. Achanta, A. Fox, and M. Skolnick, \emph{Interfacing spins in an
  ingaas quantum dot to a semiconductor waveguide circuit using emitted
  photons}, Phys. Rev. Lett. 110 (2013), p. 037402.

\bibitem{reiserer2014quantum}
A. Reiserer, N. Kalb, G. Rempe, and S. Ritter, \emph{A quantum gate between a
  flying optical photon and a single trapped atom}, Nature 508 (2014), pp.
  237--240.

\bibitem{tiecke2014nanophotonic}
T. Tiecke, J. Thompson, N. de  Leon, L. Liu, V. Vuleti{\'c}, and M. Lukin,
  \emph{Nanophotonic quantum phase switch with a single atom}, Nature 508
  (2014), pp. 241--244.

\bibitem{stern2014flux}
M. Stern, G. Catelani, Y. Kubo, C. Grezes, A. Bienfait, D. Vion, D. Esteve, and
  P. Bertet, \emph{Flux qubits with long coherence times for hybrid quantum
  circuits}, Phys. Rev. Lett. 113 (2014), p. 123601.

\bibitem{zhu2011coherent}
X. Zhu, S. Saito, A. Kemp, K. Kakuyanagi, S.i. Karimoto, H. Nakano, W.J. Munro,
  Y. Tokura, M.S. Everitt, K. Nemoto, \emph{et~al.}, \emph{Coherent coupling of
  a superconducting flux qubit to an electron spin ensemble in diamond}, Nature
  478 (2011), pp. 221--224.

\bibitem{qiu2014coupling}
Y. Qiu, W. Xiong, L. Tian, and J.Q. You, \emph{Coupling spin ensembles via
  superconducting flux qubits}, Phys. Rev. A 89 (2014), p. 042321.

\bibitem{lu2013quantum}
X.Y. L{\"u}, Z.L. Xiang, W. Cui, J.Q. You, and F. Nori, \emph{Quantum memory
  using a hybrid circuit with flux qubits and nitrogen-vacancy centers}, Phys.
  Rev. A 88 (2013), p. 012329.

\bibitem{rabl2006hybrid}
P. Rabl, D. DeMille, J.M. Doyle, M.D. Lukin, R.J. Schoelkopf, and P. Zoller,
  \emph{Hybrid quantum processors: molecular ensembles as quantum memory for
  solid state circuits}, Phys. Rev. Lett. 97 (2006), p. 033003.

\bibitem{neeley2009emulation}
M. Neeley, M. Ansmann, R.C. Bialczak, M. Hofheinz, E. Lucero, A.D. O'Connell,
  D. Sank, H. Wang, J. Wenner, A.N. Cleland, R.G. Michael, and J.M. Martinis,
  \emph{Emulation of a quantum spin with a superconducting phase qudit},
  Science 325 (2009), pp. 722--725.

\bibitem{smith2013quantum}
A. Smith, B.E. Anderson, H. Sosa-Martinez, C.A. Riofr{\'\i}o, I.H. Deutsch, and
  P.S. Jessen, \emph{Quantum control in the cs 6 s 1/2 ground manifold using
  radio-frequency and microwave magnetic fields}, Phys. Rev. Lett. 111 (2013),
  p. 170502.

\bibitem{bent2015experimental}
N. Bent, H. Qassim, A.A. Tahir, D. Sych, G. Leuchs, L.L. S{\'a}nchez-Soto, E.
  Karimi, and R.W. Boyd, \emph{Experimental realization of quantum tomography
  of photonic qudits via symmetric informationally complete positive
  operator-valued measures}, Phys. Rev. X 5 (2015), p. 041006.

\bibitem{lima2011experimental}
G. Lima, L. Neves, R. Guzm{\'a}n, E.S. G{\'o}mez, W.A.T. Nogueira, A. Delgado,
  A. Vargas, and C. Saavedra, \emph{Experimental quantum tomography of photonic
  qudits via mutually unbiased basis}, Opt. Express 19 (2011), pp. 3542--3552.

\bibitem{jaynes1963comparison}
E.T. Jaynes and F.W. Cummings, \emph{Comparison of quantum and semiclassical
  radiation theories with application to the beam maser}, Proceedings of the
  IEEE 51 (1963), pp. 89--109.

\bibitem{shore1993jaynes}
B.W. Shore and P.L. Knight, \emph{The {J}aynes-{C}ummings model}, Journal of
  Modern Optics 40 (1993), pp. 1195--1238.

\bibitem{blais2004cavity}
A. Blais, R.S. Huang, A. Wallraff, S.M. Girvin, and R.J. Schoelkopf,
  \emph{Cavity quantum electrodynamics for superconducting electrical circuits:
  An architecture for quantum computation}, Phys. Rev. A 69 (2004), p. 062320.

\bibitem{wallquist2009hybrid}
M. Wallquist, K. Hammerer, P. Rabl, M. Lukin, and P. Zoller, \emph{Hybrid
  quantum devices and quantum engineering}, Physica Scripta 2009 (2009), p.
  014001.

\bibitem{groblacher2009observation}
S. Gr{\"o}blacher, K. Hammerer, M.R. Vanner, and M. Aspelmeyer,
  \emph{Observation of strong coupling between a micromechanical resonator and
  an optical cavity field}, Nature 460 (2009), pp. 724--727.

\bibitem{mischuck2013qudit}
B. Mischuck and K. M{\o}lmer, \emph{Qudit quantum computation in the
  jaynes-cummings model}, Phys. Rev. A 87 (2013), p. 022341.

\bibitem{blume02a}
R. Blume-Kohout, C.M. Caves, and I.H. Deutsch, \emph{Climbing mount scalable:
  Physical resource requirements for a scalable quantum computer}, Found.~Phys.
  32 (2002), pp. 1641--1670, arXiv:quant-ph/0204157.

\end{thebibliography}
\end{document}